\documentclass[prl, twocolumn, superscriptaddress, floatfix, tighten]{revtex4-2}

\usepackage{xcolor}
\usepackage{amsmath}
\usepackage{amssymb}
\usepackage{bm}      	
\usepackage{graphicx}
\usepackage[unicode=true,pdfusetitle,
 bookmarks=false,bookmarksnumbered=false,bookmarksopen=false,
 breaklinks=false,pdfborder={0 0 1},backref=false,colorlinks=true,urlcolor=blue,citecolor=blue,linkcolor=black]
 {hyperref}
\usepackage{subfigure}
\usepackage{textcomp}
\usepackage{microtype} 
\usepackage[nameinlink]{cleveref}

\newcommand{\bsub}{\begin{subequations}}
\newcommand{\esub}{\end{subequations}}

\newcommand{\vex}[1]{\bm{\mathrm{#1}}}

\crefname{figure}{Fig.}{Figs.}
\Crefname{figure}{Fig.}{Figs.}
\crefname{equation}{Eq.}{Eqs.}
\Crefname{equation}{Eq.}{Eqs.}

\newif\ifpanelparens
\panelparensfalse

\newcommand{\panel}[1]{\ifpanelparens (#1)\else \textbf{#1}\fi}

\begin{document}

\title{Twisted Trilayer Graphene, Quasiperiodic Superconductor} 
\def\rice{Department of Physics and Astronomy, Rice University, Houston, Texas
77005, USA}
\def\rcqm{Rice Center for Quantum Materials, Rice University, Houston, Texas
77005, USA}
\author{Xinghai Zhang}\affiliation{\rice}
\author{Ziyan Zhu}
\affiliation{Department of Physics, Boston College, 140 Commonwealth Avenue
Chestnut Hill, MA 02467, USA
}
\author{Justin H. Wilson}
\affiliation{Department of Physics and Astronomy, Louisiana State University, Baton Rouge, LA 70803, USA}
\affiliation{Center for Computation and Technology, Louisiana State University, Baton Rouge, LA 70803, USA}
\author{Matthew S. Foster}\affiliation{\rice}\affiliation{\rcqm}
\date{\today}

\maketitle


{\bf 
Twisted multilayer moir\'e materials are generically quasiperiodic on the moir\'e scale due to the interference of different misaligned moir\'e periodicities.
Spatial inhomogeneities such as these 
can be 
detrimental to superconductivity~\cite{Zhang2022,Ticea2024}; nonetheless, superconductivity has been observed in quasiperiodic twisted trilayer graphene (TTG)~\cite{Uri2023}.
Here, we systematically study the superconducting properties of TTG. 
We reveal
that an interplay between quasiperiodicity and topology drives TTG into a critical regime,
enabling it to host superconductivity with rigid phase stiffness for a wide range of twist angles, rather than at a fine-tuned value. 
The criticality in the normal state is due to the Dirac fermions coupled by quasiperiodic tunneling \emph{simulating}
3D topological superconductor surface states
\cite{Zhang2025}.
This critical-metal regime is 
marked by multifractal 
wave functions across the spectrum and scale-invariant transport reminiscent of the integer quantum Hall plateau transition. 
We demonstrate this with
large-scale wave function and Kubo conductivity calculations. 
These observations lead to a clear experimental implication: 
stronger interlayer coupling in TTG further stabilizes both the criticality and superconductivity, allowing superconductivity to be seen across a wider range of angles 
with experimentally accessible pressures.}

Twisted multilayer graphene has recently emerged as a powerful platform for exploring strongly correlated electronic phases.
Interference between multiple moiré patterns generates additional long-wavelength scales beyond those in bilayer systems. 
The secondary moiré-of-moiré scale~\cite{zhu20202mechanical} can be substantially larger than individual moir\'e patterns even when slightly misaligned, 
and it is generally quasiperiodic (incommensurate) in the continuum limit.
This additional structural degree of freedom leads to a variety of electronic behaviors~\cite{turkel2022,craig2024,hoke2024imaging,park2025} and creates a landscape distinct from that of the bilayer case~\cite{Zhu2020,Devakul2023,Mao2023,Guerci2024,Yang2024}. 
Nonetheless, recent experiments on twisted trilayer graphene (TTG) reveal superconductivity~\cite{Zhang2021,turkel2022,Uri2023,Xia2025a,hoke2025linking}, anomalous Hall responses~\cite{Xia2025,Xia2025a}, and correlated insulating states~\cite{Zhang2021,hoke2025linking,Xie2025}.
These observations suggest TTG is not a simple perturbation of TBG, but a distinct platform for new correlated phases.

Quasiperiodicity in low dimensions is usually expected to suppress transport by driving electronic states towards 
Anderson
localization \cite{Sokoloff1985}. 
This raises a fundamental question: how can robust superconductivity appear in a system whose underlying single-particle electronic wave functions are, naively, strongly inhomogeneous or even localized? 
In ordinary two-dimensional (2D) systems, increasing spatial inhomogeneity predominantly suppresses phase coherence and reduces the superfluid stiffness, often destroying superconductivity, rather than promoting it~\cite{Trivedi1996, Fan2021, Ticea2024}. 

The key is that graphene-based moir\'e materials differ from conventional 2D systems. 
Their low-energy Dirac fermions 
are coupled between layers by
effective long-wavelength non-Abelian gauge 
potentials. 
This places TTG in a class analogous to topological surface states, where topology can protect the system from Anderson localization and stabilize critical transport even in the presence of strong inhomogeneity~\cite{Sbierski2020, Karcher2021, Zhang2025}.
As a result, quasiperiodic TTG can enter a 
metallic regime in which 
\emph{all} wave functions
across a broad energy window
become 
quantum critical (``multifractal'' \cite{Evers2008})
and transport acquires a scale-invariant character reminiscent of the integer quantum Hall plateau transition \cite{Sbierski2020}. 
This ``critical metal'' is neither ballistic nor localized: with conductivity $\sim e^2/h$ over a wide range of energies. 
Such critical wave functions are known to enhance pairing~\cite{Feigelman2007, Tezuka2010, Fan2021, Zhang2022, Zhang2025}, providing a mechanism by which quasiperiodicity can stabilize robust superconductivity in TTG.

\begin{figure*}[ht!]
    \centering
    \includegraphics[width=0.98\textwidth]{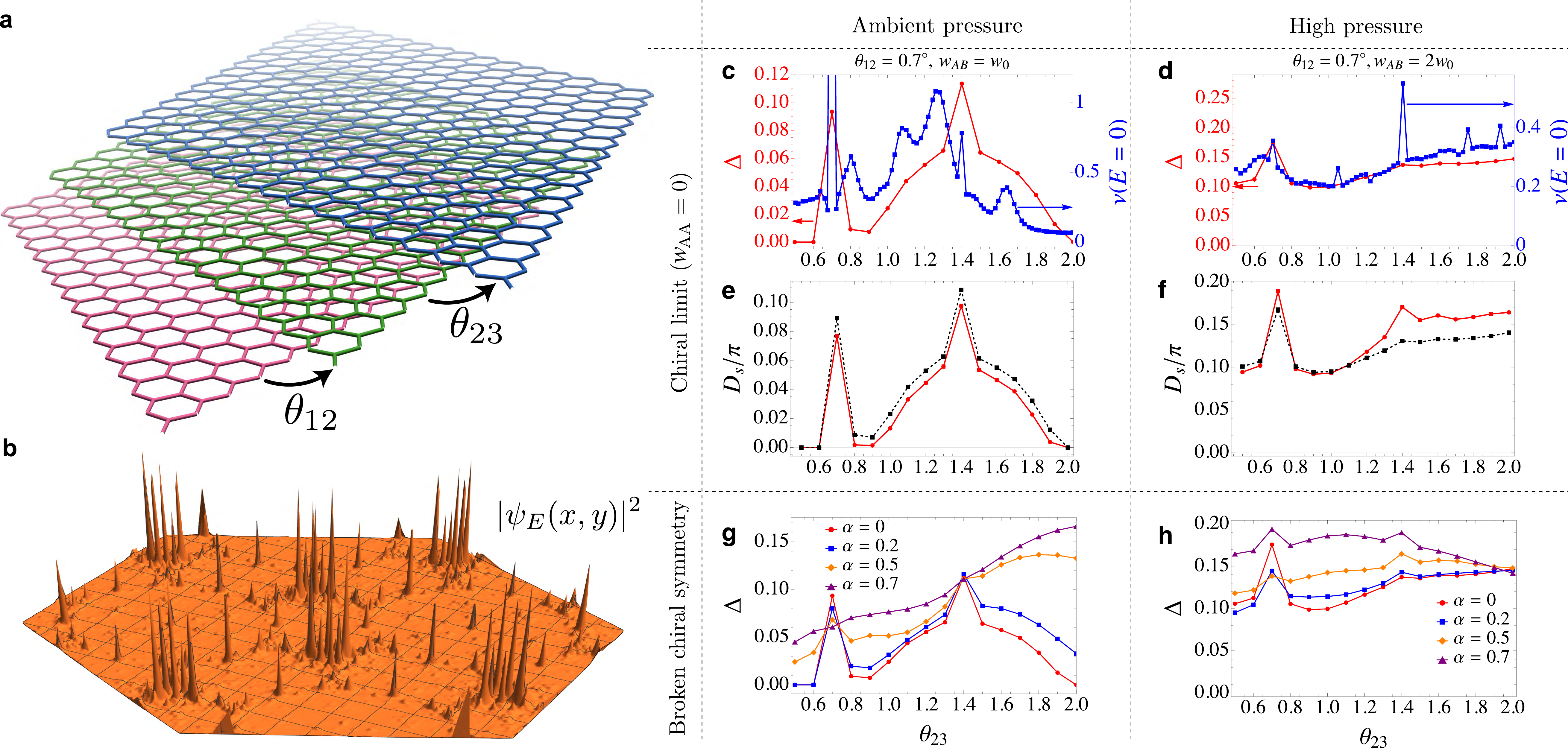}
    \caption{{\bf Fractal wave functions and robust superconductivity in TTG.}
    \panel{a}, Illustration of helical twisted trilayer graphene with two independent twist angles 
    $\theta_{12}$ and $\theta_{23}$.
    \panel{b}, A typical 
    quantum-critical, normal-state single-particle
    wave function in quasiperiodic TTG ($\theta_{12} =0.5^\circ$ and $\theta_{23}=0.925^\circ$).
    The self-similar structure in position space is induced by the quasiperiodicity.
    \panel{c}, 
    Spatial average of the
    superconductivity order parameter $\Delta$ and 
    density of states near the Fermi energy $\nu(E = 0)$ 
    in
    the chiral limit \cite{Tarnopolsky2019} with $w_\mathrm{AB}=w_0=110 $ meV; 
    \panel{d}, Same as \panel{c}, but with doubled interlayer coupling $w_\mathrm{AB}=2 w_0=220$ meV.
    \panel{e} and \panel{f}, Superfluid stiffness $D_s/\pi$ (red) in the chiral limit with $w_\mathrm{AB}=w_0$ and $w_\mathrm{AB}=2w_0$, respectively. 
    The black dashed line shows $D_s/\pi  = 3\Delta/\pi $, which is the result for an \emph{ideal} Dirac superconductor without moir\'e potentials \cite{Kopnin2008,Uchoa2009,Kopnin2009,Zhang2025}.  
    \panel{g} and \panel{h}, Superconducting order parameter 
    for the realistic TTG model with broken
    chiral symmetry 
    ($\alpha\ne 0$, see text) 
    with
    $w_\mathrm{AB}=w_0$ and $w_\mathrm{AB}=2w_0$, respectively. 
    $\Delta$ and $D_s/\pi$ 
    are given in units 
    of $v_F k_{\theta_{12}} \approx 124$ meV and the attractive interaction is set to be $U=4 v_F k_{\theta_{12}} =496$ meV. 
    The linear system size is $L = 77$ for \panel{c}--\panel{h},
    while the polynomial order $N_\mathcal{C}$ utilized in the kernel polynomial method is $2048$ for $\Delta$ and $1024$ for $D_s$ calculations
    \cite{SM}.
    }
    \label{fig:SC-ttg}
\end{figure*}

Here we demonstrate quasiperiodicity-enhanced superconductivity in TTG.
Using large-scale continuum-model simulations, we show that generic incommensurate twist-angle pairs yield a superconducting phase with sizable superfluid stiffness across a broad range of angles.
We attribute this behavior to the presence of quantum-critical electronic states in the normal phase, as seen in multifractal wave function diagnostics and Kubo conductivity.
We further show that increasing the interlayer coupling---achievable via hydrostatic pressure~\cite{Carr2018,yankowitz2019,Wang2025}---stabilizes the criticality and strengthens superconductivity.
Together, these results establish TTG as a platform where quasiperiodicity and electronic topology conspire to stabilize superconductivity beyond a narrow, fine-tuned flat-band (magic-angle) regime.

\section*{Superconductivity in TTG}

We focus on helical TTG with independent tunable twist angles $\theta_{12}$ and $\theta_{23}$ [\cref{fig:SC-ttg}\panel{a}]. 
The quasiperiodicity is governed by the competition of the two moir\'e scales 
$L_{12} \sim {\sqrt{3} a}/{\theta_{12}}$ 
and 
$L_{23}\sim {\sqrt{3} a}/{\theta_{23}}$ 
(small-angle limit; $a\approx 1.42$~\AA{}).
In this section we establish the presence and robustness of superconductivity in the resulting inhomogeneous landscape. 
In the following section, we show that it originates from the quantum-critical normal-state electrons.

The low-energy physics of TTG in valley $K$
is described by the Hamiltonian \cite{BM2011, Zhu2020},
\begin{equation}
\begin{aligned}
    H_{K} 
    & =
    \int
    d^2\vex{r}
    \,\,
    \psi_{K\mathbf{r}}^{\dagger}
    \,
    h_{K}\left(\mathbf{r}\right)
    \,
    \psi_{K\mathbf{r}}\,,
\\
    h_{K}\left(\mathbf{r}\right)
    & =
    \left[
        -i\boldsymbol{\sigma}\cdot\nabla
        +
        V_{12}\left(\mathbf{r}\right)
        +
        V_{23}\left(\mathbf{r}\right)
    \right]\,,
\end{aligned}
\end{equation}
where $\psi_{K\mathbf{r}}$ is a six-component 
electronic
spinor with sublattice
($\sigma=1,2$) and layer indices ($1,2,3$). 
$V_{12}\left(\mathbf{r}\right)$ and $V_{23}\left(\mathbf{r}\right)$ are the 
interlayer tunneling potentials forming the
top and bottom moir\'e patterns.
$V_{12}(\mathbf{r})$ takes the form
\begin{equation}\label{eq:V12Def}
\begin{aligned}
    V_{12}\left(\mathbf{r}\right)  
    & =
    w_{\mathrm{AB}} 
    \left[
        U_{1}\left(\mathbf{r}\right)t^{+}\sigma^{+}
        +
        U_{-1}\left(\mathbf{r}\right)t^{+}\sigma^{-}
    \right]
\\
    &\phantom{=}+ 
    w_{\mathrm{AA}} 
    \,
    U_{0}\left(\mathbf{r}\right)t^{+}
    + 
    \mathrm{H.c.}\,.
    \end{aligned}
\end{equation}
Here 
$
U_{n}
\left(\mathbf{r}\right)
=
e^{i\mathbf{q}_{1}^{(12)}\cdot\mathbf{r}}
+
e^{-in\phi}e^{i\mathbf{q}_{2}^{(12)}\cdot\mathbf{r}}
+
e^{in\phi}e^{i\mathbf{q}_{3}^{(12)}\cdot\mathbf{r}}
$ 
with $n=0,\pm 1$, and $t^{\pm}=\frac{1}{2}\left(t^{1}\pm i t^{2}\right)$
are $3 \times 3$ matrices coupling together layers $1$ and $2$.
In \cref{eq:V12Def},
$w_\mathrm{AA}$ and $w_\mathrm{AB}$ are the interlayer tunneling strengths 
in the AA and AB (Bernal) stacking regions, respectively. 
The tunneling in the AA region is smaller than that in the AB region due to lattice relaxation effects~\cite{koshino2017,carr2018relaxation}, 
which can be quantified by the dimensionless parameter $\alpha \equiv w_\mathrm{AA}/w_\mathrm{AB} \approx 0.7$.
The moir\'e potential between layers $1$ and $2$ is characterized by the reciprocal lattice vectors 
$\mathbf{q}_{1}^{(12)}=k_{\theta_{12}}\left(1,0\right)$
and 
$\mathbf{q}_{2,3}^{(12)}=k_{\theta_{12}}\left(-1/2,\pm\sqrt{3}/2\right)$,
with $k_{\theta_{12}}=2K\sin(\theta_{12}/2)$.
The interlayer coupling $V_{23}$ 
takes a similar form with $\mathbf{q}_{i}^{(12)}$ replaced
by $\mathbf{q}_{i}^{(23)}$. 

In this paper, we mostly employ a
collinear approximation and assume that 
$\mathbf{q}_{i}^{(23)} = \mathbf{q}_{i}^{(12)} \sin\frac{\theta_{23}}{2} / \sin\frac{\theta_{12}}{2}$, in order to access much larger system sizes than the previous work based on a non-collinear model~\cite{Zhu2020}. 
In other words, we assume 
that the two adjacent moir\'e lattices are rotationally aligned 
(but with quasiperiodic periods).
This is justified at small twist angles where the misalignment at the moir\'e-of-moir\'e scale~\cite{zhu20202mechanical} is parametrically weak (we check that collinear and non-collinear models agree qualitatively, see below). 

The chiral limit 
with $\alpha = 0$
is believed to capture the key features of the twisted 
multilayer
graphene and is a good starting point for theoretical study \cite{Tarnopolsky2019}. In this limit, $V_{12}+V_{23}$ is exactly a non-Abelian gauge potential and the system mimics the topological \emph{surface states} of a 3D topological superconductor, 
explained below.

For simplicity, we focus on the simplest conventional scenario for superconductivity: intervalley $s$-wave pairing.
(More exotic pairing scenarios are discussed in the conclusion.)
We consider a local attractive interaction
\begin{equation}
    H_{\mathsf{int}}
    =
    -
    U
    \int
    d^2\vex{r}
    \,\,
    \psi_{\mathbf{r}\uparrow}^{\dagger}
    \,
    \psi_{\mathbf{r}\uparrow}
    \,
    \psi_{\mathbf{r}\downarrow}^{\dagger}
    \,
    \psi_{\mathbf{r}\downarrow}\,,
\label{eq:HU}
\end{equation}
and treat it within self-consistent Bogoliubov-de Gennes (BdG) theory to obtain the local superconducting pairing amplitude $\Delta_\mathbf{r}$ and density $n_{\mathbf{r}}$.
The BdG Hamiltonian is 
\begin{equation}
    H_{\mathsf{BdG}}
    =
    \frac{1}{2}
    \int
    d^2\vex{r}
    \,\,
    \chi_{\mathbf{r}}^{\dagger}
    \,
    h_{\mathsf{BdG}}
    \,
    \left(\mathbf{r}\right)
    \chi_{\mathbf{r}}\,,
\end{equation}
where quasiparticles are encoded in the composite spinor
$\chi=[\psi_{K\mathbf{r}},i s_{2}(\psi_{-K\mathbf{r}}^{\dagger})^{\mathsf{T}}]^{\mathsf{T}},$
the valleys are labeled by $\pm K$,
and the BdG Hamiltonian is 
\begin{equation}
    h_{\mathsf{BdG}}\left(\mathbf{r}\right)
    =
    \left[
        h_{K}\left(\mathbf{r}\right) - \frac{U}{2} n_{\mathbf{r}}
    \right]
    \mu_{3}
    +
    \Delta_\mathbf{r} \, \mu_{1}.
\end{equation}
In the above, $s_i$ and $\mu_j$ respectively denote Pauli matrices acting on spin-1/2 and particle-hole spaces;
$-U \, n_{\mathbf{r}}/2$ is the Hartree shift. 
The ground-state superconductivity is evaluated self-consistently via the kernel polynomial method (KPM);
details of the computational methodology can be found in Refs.~\cite{Zhang2025,SM}.
In what follows, we focus on the case of half-filling.

We begin in the chiral limit with standard~\cite{BM2011,koshino2017} interlayer couplings $w_\mathrm{AA}=0$ and $w_\mathrm{AB}=110$~meV. 
The spatially averaged pairing amplitude $\Delta$ and zero-energy density of states (DOS) $\nu(E=0)$ 
are
plotted versus twist angle in 
\Cref{fig:SC-ttg}\panel{c}.
Both the DOS and the pairing amplitude exhibit sharp maxima at commensurate configurations where $\theta_{23}$ is an integer multiple of $\theta_{12}$.
Away from these special angles, however, the DOS and $\Delta$ do not track each other, implying that superconductivity is not simply controlled by flat-band physics. 
\Cref{fig:SC-ttg}\panel{e} shows the superfluid stiffness of quasiperiodic TTG.
The superfluid stiffness $D_s$ closely follows the analytical prediction expected for 
an \emph{ideal} Dirac superconductor (without moir\'e potentials),
$D_{s}/\pi=3\Delta/\pi$ \cite{Kopnin2008,Uchoa2009,Kopnin2009,Zhang2025}.
This behavior shows that the superconductivity in TTG is robust against phase fluctuations and that quasiperiodicity does not degrade phase coherence. 
The reason for this turns out to be the topology in the TTG Hamiltonian in the chiral limit, 
which simulates topological surface states. 
The normal state is critical, with a conductivity
pinned to a universal value $\sigma_{n}=3e^{2}/{\pi h}$ reflected in the superfluid stiffness
\cite{Zhang2025}.
This finite and angle-insensitive conductivity prevents the suppression of 
$D_s$ 
that would normally arise from strong spatial inhomogeneity \cite{Trivedi1996}, allowing superconductivity to remain robust across a broad range of twist angles.

All of this physics is enhanced and changes qualitatively when we simulate applied pressure by doubling the interlayer coupling to $w_\mathrm{AA}=0$, $w_\mathrm{AB}=220$ meV, as shown in \cref{fig:SC-ttg}\panel{d}.
Such enhanced couplings are experimentally accessible using 
realistic
hydrostatic pressures~\cite{Carr2018,Wang2025,yankowitz2019}.
Commensurate configurations still produce noticeable peaks in the average $\Delta$, but the pairing amplitude becomes far more uniform across twist-angle space. This robustness 
implies
that increasing $w_\mathrm{AB}$ removes the angle sensitivity and thus stabilizes superconductivity across twist angles. 
Similar behavior is seen with the superfluid stiffness in \cref{fig:SC-ttg}\panel{f}, which still roughly follows $D_{s}/\pi=3\Delta/\pi$ implying it too is angle-robust.

In realistic samples, 
although
Bernal (AB) stackings have a smaller interlayer spacing due to lattice relaxation and $w_\mathrm{AB} > w_\mathrm{AA}$~\cite{koshino2017,carr2018relaxation}, $w_\mathrm{AA}$ never vanishes. 
The $w_\mathrm{AA}$ term breaks both the approximate chiral symmetry \cite{Tarnopolsky2019} of the continuum model 
and the link to topological surfaces, 
but it does not gap the Dirac points or generate a \emph{Chern mass}. 
As we explain below, the finite-energy wave functions of the chiral model in the normal state show critical fluctuations consistent with the integer quantum Hall plateau transition. In the absence of nonzero average Berry curvature, this criticality is robust \cite{Altland2024}.
\Cref{fig:SC-ttg}\panel{g,h} show the pairing amplitude of TTG with broken chiral symmetry ($\alpha\neq0$) for $w_\mathrm{AB}=w_0$ and $w_\mathrm{AB}=2w_0$.
At
commensurate angles, superconductivity is suppressed or unchanged by the chiral symmetry-breaking term (except for $\alpha=0.7$ and $w_\mathrm{AB}=2w_0$) compared to the chiral limit. 
By contrast, for incommensurate angles, where quasiperiodicity is strongest, the symmetry-breaking term 
\emph{further enhances} the pairing amplitude.
This behavior 
demonstrates
that the superconductivity supported by the quasiperiodic normal state is not tied to fine-tuned chiral symmetry. Rather, the mechanism that stabilizes pairing in TTG remains effective even when AA tunneling is included at realistic strengths.

The robust superfluid stiffness in the presence of quasiperiodicity indicates that the normal state does not Anderson-localize but remains in a critical conducting regime.
We now show TTG effectively simulates a class-AIII topological surface that exhibits spectrum-wide quantum criticality, yielding multifractal wave functions and near-universal longitudinal conductivity.

\begin{figure}[t!]
    \centering
    \includegraphics[width=0.99\columnwidth]{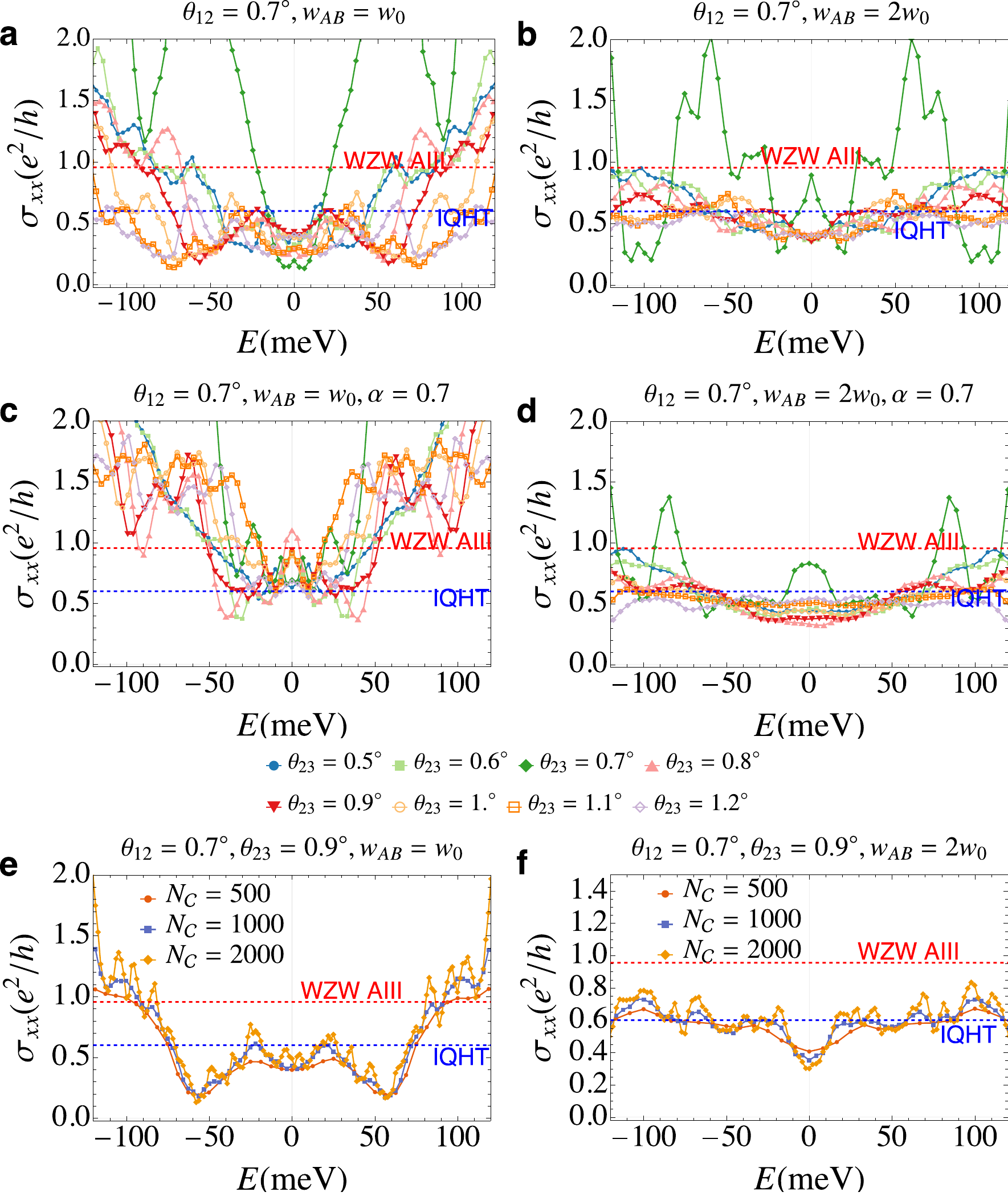}
    \caption{{\bf Normal-state conductivity in TTG.} 
    \panel{a--b}, 
    Conductivity $\sigma_{xx}$ 
    of TTG in the chiral limit 
    versus eigenstate energy $E$
    with $w_\mathrm{AB}=w_0$ and $w_\mathrm{AB}=2w_0$, respectively. 
    \panel{c--d}, 
    Including realistic
    chiral symmetry breaking 
    $\alpha=0.7$. 
    \panel{e--f}, 
    Conductivity evaluated with 
    different KPM polynomial orders
    $N_\mathcal{C}$. 
    The red and blue dashed lines show 
    reference values for
    critical states predicted to occur in class AIII \cite{Ludwig1994,Sbierski2020}:
    WZW [zero energy, $3 e^2/(\pi h)$] and IQHT [finite energy, $\approx 0.6 e^2/h$].
    The linear system size $L = 308$.}
    \label{fig:conductivity}
\end{figure}

\section*{Quasiperiodic criticality}

Surfaces of 3D bulk topological phases~\cite{Schnyder2008,Kitaev2009} exhibit massless two-dimensional (2D) Dirac electrons with \emph{unique} properties not usually realizable in any crystalline or quasiperiodic 2D system with local hopping \cite{Redlich1984,Semenoff1984}. 
However, they can be \emph{effectively simulated} by such a system if intervalley scattering can be neglected. This is the case with small twist-angle multilayer graphene \cite{Zhang2025}. 

In the chiral limit and neglecting intervalley coupling, the Dirac electrons in the normal state of TTG simulate the surface of a class-AIII topological phase \cite{Schnyder2008,Sbierski2020,Karcher2021}. 
This surface is predicted to exhibit zero-energy states that are topologically protected from Anderson localization and that possess a universal longitudinal conductivity \cite{Ludwig1994}. 
Recently it was hypothesized and numerically demonstrated that the finite-energy states of a class-AIII surface with symmetry-preserving 
disorder exhibit a remarkable ``spectrum-wide quantum criticality'' (SWQC) \cite{Karcher2021}. 
Disorder and topology induce a ``multifractal metal''---a whole tower of robust finite-energy states~\cite{Sbierski2020,Altland2024} with universal statistics consistent with the integer quantum Hall plateau transition (IQHT), itself a 2D topological quantum phase transition.
By contrast, in the quantum Hall effect critical IQHT states occur only at isolated energies.

\begin{figure}
    \centering
    \includegraphics[width=0.98\columnwidth]{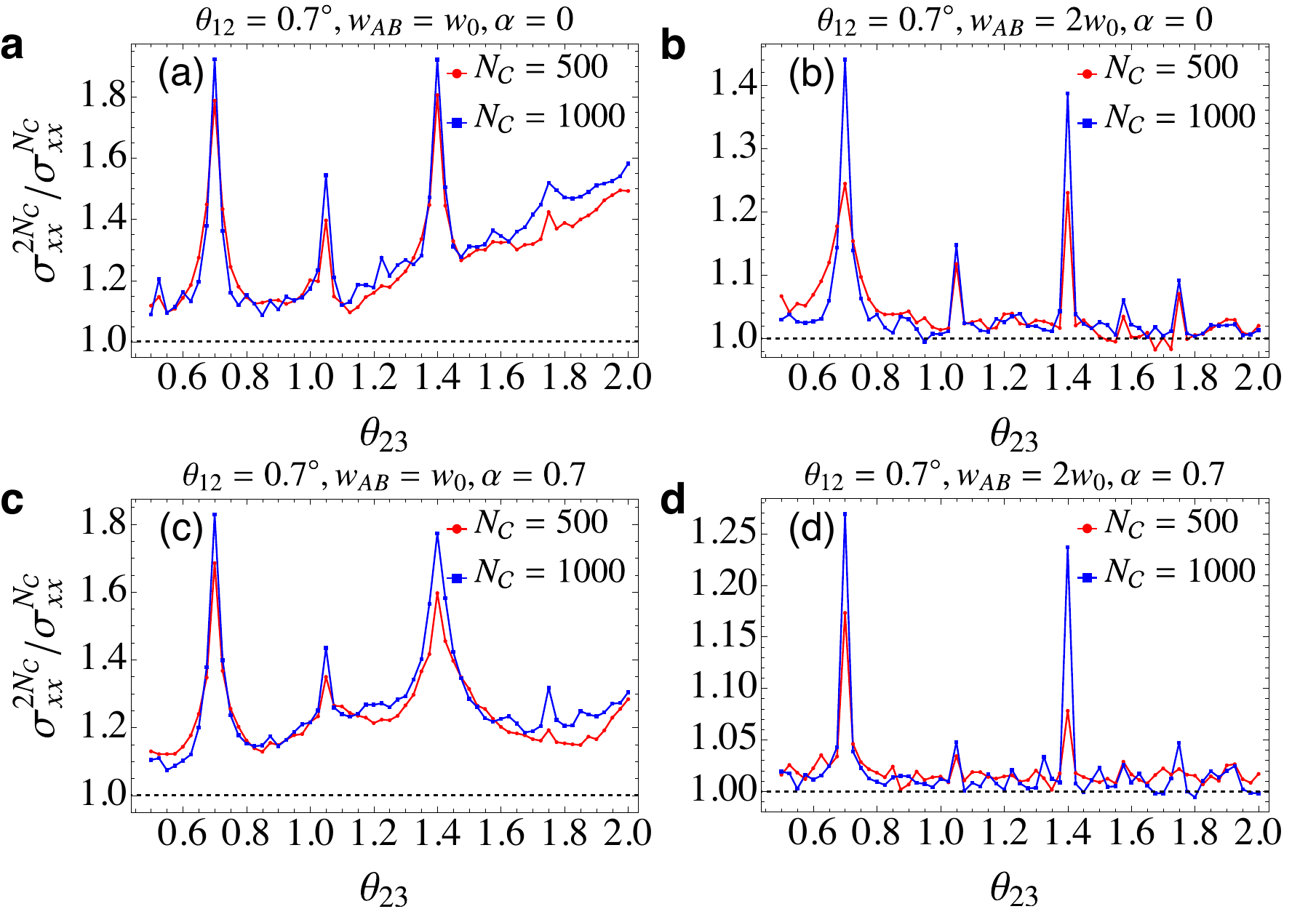}
    \caption{{\bf The ratio of normal-state conductivity evaluated with $2N_\mathcal{C}$ and $N_\mathcal{C}$.} 
    \panel{a-b}, The ratio $\sigma_{xx}^{2N_\mathcal{C}}/\sigma_{xx}^{N_\mathcal{C}}$ of TTG in the chiral limit with $w_\mathrm{AB}=w_0$  and $w_\mathrm{AB}=2w_0$, respectively. 
    \panel{c-d}, The ratio for TTG with chiral symmetry breaking. 
    The ratio is expected to be $\approx 1$ (dashed line) for critical states, and deviates from $1$ for ballistic transport. 
    The peaks in the ratio correspond to  
    commensurate twist angles.
   Results are averaged over the energy window $v_F k_{\theta_{12}} (-1,1)$
    and $L = 308$.
    }
    \label{fig:sigmaxx-th07-ratio}
\end{figure}

Quasiperiodicity can play the same role as symmetry-preserving disorder.
In TTG, the AIII-like symmetry is approximately preserved by the suppression of intervalley scattering and the nature of tunneling between layers, allowing it to realize the SWQC scenario involving IQHT states across the energy spectrum.
A key attribute of this phenomenology is that the normal-state conductivity should be of order $e^2/h$ for a wide swath of eigenstate energies $E$.
The normal-state conductivity (in units of $e^2/h$) is given by the Kubo formula,
\begin{equation}
    \sigma_{xx}(E) 
    =
    \frac{2\pi^2}{L^2}
    \mathsf{Tr}
    \left[
        j_x \, \delta(E - h_K) \, j_x \, \delta(E - h_K)
    \right] \,.
    \label{eq:sigmaxx}
\end{equation}
Here $j_x$ is the current operator in $x$-direction
and $L$ is the linear system size.
We calculate the Kubo conductivity 
using
KPM~\cite{weibe2006_kpm,SM}, and access system sizes comparable to experimental samples. 
The normal-state conductivity $\sigma_{xx}(E)$ of TTG is shown in 
\cref{fig:conductivity}.
In the chiral limit, the calculated TTG
conductivity fluctuates around the 
average IQHT
critical conductivity $\approx 0.6 e^2/h$ \cite{Jovanovic1998, Schweitzer2005}
for incommensurate twist angles [\cref{fig:conductivity}\panel{a}],  
consistent with the topological SWQC scenario.
At $\theta_{12}=\theta_{23} = 0.7^\circ$, the
collinear model
is commensurate 
with a finite moir\'e period,
and the conductivity 
well exceeds
this critical value. 
This is consistent with ballistic transport
broadened by the KPM method (see below).
Increasing the interlayer tunneling strength to $w_\mathrm{AB}=2w_0$ [\cref{fig:conductivity}\panel{b}], the 
chiral TTG model 
conductivity
locks closely to the IQHT value
with weaker fluctuations 
over a
wider range of energy, 
consistent with the superconducting behavior seen in \cref{fig:SC-ttg}\panel{d,f}.

Incorporating realistic 
chiral symmetry breaking 
($\alpha\approx 0.7$) increases the conductivity in 
quasiperiodic TTG with $w_\mathrm{AB}=w_0$ [\cref{fig:conductivity}\panel{c}], and weakens the criticality.
By contrast,
conductivity is 
stabilized near the IQHT value when the interlayer tunneling is doubled, 
even with broken chiral symmetry [\cref{fig:conductivity}\panel{d}].
These results
show that 
TTG at normal pressure ($w_\mathrm{AB}=w_0$) 
resides in a crossover regime 
featuring both ballistic and critical states, 
where  
wave functions 
are
sensitive to microscopic details, especially 
chiral symmetry breaking.
The possibility of ballistic states despite broken translational symmetry is a well-known feature of quasiperiodic systems \cite{Sokoloff1985}. 
\cref{fig:conductivity}\panel{d} shows that the topological criticality
is uniformly stabilized with increasing interlayer tunneling, as achievable via applied pressure, even without chiral symmetry---again 
complimentary to the enhanced 
superconducting behavior seen in \cref{fig:SC-ttg}\panel{h}.

\begin{figure*}[ht!]
    \centering
    \includegraphics[width=0.7\textwidth]{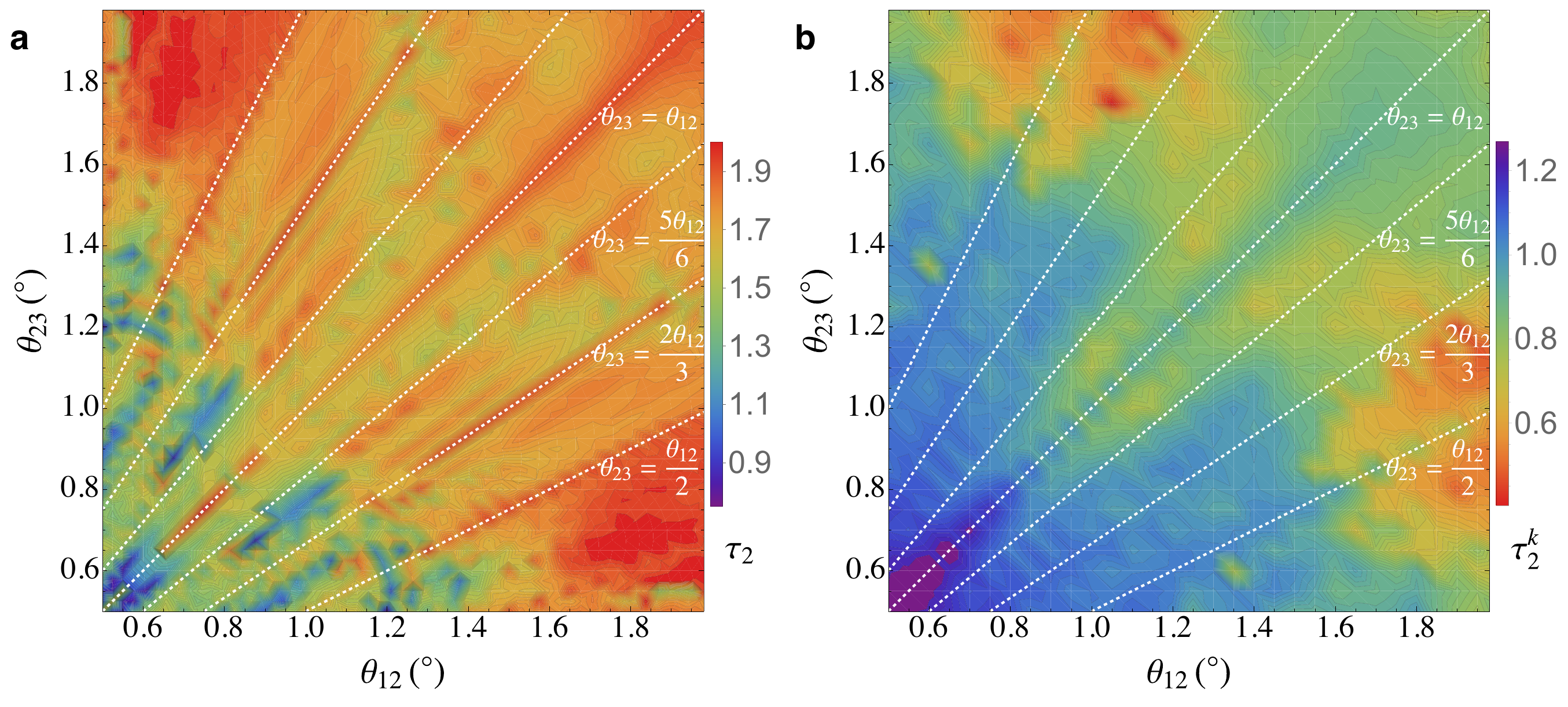}
    \caption{ {\bf Multifractal dimension in collinear and non-collinear model for TTG.} Here we show results in the chiral limit with $w_\mathrm{AB}=w_0$. 
    \panel{a}, Real-space multifractal dimension $\tau_2$ 
    for the $E = 0$ wave function in TTG evaluated with the collinear model. 
    \panel{b}, Momentum-space multifractal dimension $\tau_2^k$ in TTG evaluated with the non-collinear model. 
    Small $\tau_2$ and large $\tau_2^k$ indicate a quantum-critical state.
    The multifractal dimensions in both models reflect the same criticality behavior in the normal state of TTG.
}
    \label{fig:mfc-diagram}
\end{figure*}

To distinguish between ballistic and critical or diffusive regimes in quasiperiodic systems, KPM conductivity calculations provide a natural diagnostic.
When using KPM on \cref{eq:sigmaxx}, 
the
conductivity is naturally regulated by a lifetime $\tau_\mathrm{KPM} \sim N_\mathcal{C}$,
where $N_\mathcal{C}$ is the order of the polynomials employed.
In critical or diffusive regimes, the system develops a physical lifetime $\tau_\mathrm{phys}$ 
due to elastic impurity scattering,
which is independent of $N_\mathcal{C}$. 
When
$\tau_\mathrm{KPM} \gtrsim \tau_\mathrm{phys}$ 
the conductivity will plateau with 
increasing
$N_\mathcal{C}$ while developing noisy state-to-state fluctuations for very large $N_\mathcal{C}$,
see \cref{fig:conductivity}\panel{e} and \panel{f}.
For ballistic states, however, the conductivity is 
(artificially)
determined by $\tau_\mathrm{KPM}$ 
and increases 
with $N_\mathcal{C}$ without saturating.

\Cref{fig:sigmaxx-th07-ratio} shows the ratio of $\sigma_{xx}$ evaluated with different $N_\mathcal{C}$
plotted versus twist angle.
The ratio is close to unity at incommensurate angles and peaks around the commensurate ones.
This shows that
quasiperiodic TTG is critical, while the commensurate 
angles are
ballistic.
At large $\theta_{23}$, 
quasiperiodic TTG also becomes ballistic as the third graphene layer 
effectively decouples from the other two.
With chiral symmetry breaking [\cref{fig:sigmaxx-th07-ratio}\panel{c}], the peaks are broadened, indicating weakened criticality.
Instead,
the ratio collapses to $1$ for incommensurate twist angles and shows narrow peaks at commensurate angles [\cref{fig:sigmaxx-th07-ratio}\panel{b} and \panel{d}] when interlayer coupling is doubled,
consistent with 
enhanced criticality in the conductivity, \cref{fig:conductivity}\panel{d}.

\section*{Model comparison}

In TTG, each layer’s momentum-space Dirac lattice is rotated by a different twist angle. 
As two layers are twisted relative to each other, both the magnitude and direction of the moir\'e reciprocal vectors change; the collinear model neglects the directional mismatch.
Keeping both magnitude and direction mismatch leads to a ``non-collinear'' model \cite{Zhu2020} which is more computationally intensive than the collinear model employed above.
However, for small twist angles the directional mismatch between layers is small relative to the magnitude mismatch, justifying the collinear approximation. 
Here, we show both the collinear model and the non-collinear model capture the same underlying physics. 

We first show that both models predict qualitatively the same multifractal behaviors 
for critical wave functions at zero energy $E = 0$. 
Wave function criticality is characterized by the scaling of the inverse participation ratio (IPR) $P_2 \equiv \sum_{\mathbf r} \left|\psi_\mathbf{r}\right|^4 \sim L^{-\tau_2}$, where $L$ is the linear system size and $\tau_2$ is the multifractal dimension in real space, defined by $\mathbf r$. 
We can also define a momentum-space IPR by replacing 
$\mathbf r$ with $\mathbf k$.
In the collinear model, we directly access real-space wave functions and extract the real-space exponent $\tau_2$, see \cref{fig:mfc-diagram}\panel{a}. 
For $\theta_{12} + \theta_{23} \lesssim 1.6^\circ$ and away from commensurate angles we find $\tau_2 \lesssim 1.8$, indicating multifractal, critical states.
Along the commensurate lines (white dashed lines), the system becomes periodic in the continuum limit, and $\tau_2 \approx 2$, consistent with plane-wave-like states. 
Similarly, in the strongly asymmetric limit $\theta_{12} \gg \theta_{23}$, the system effectively reduces to a weakly coupled twisted bilayer and a monolayer.
Approximate moir\'e periodicity is restored, and wave functions again become extended, consistent with ballistic transport. 

In the non-collinear model \cite{SM}, we have direct access to the momentum-space IPR, which defines the momentum-space multifractal exponent $\tau_2^k$. 
With the momentum-space IPR, the trends are inverted: plane-wave-like states correspond to small $\tau_2^k$, while strongly localized states yield $\tau_2^k \approx 2$. 
As shown in \cref{fig:mfc-diagram}\panel{b}, the non-collinear model exhibits a broad critical region with $\tau_2^k \sim 1$ in the same triangular domain of twist angles where the collinear model has real-space multifractality.  
In the large-angle, weakly coupled regimes, $\tau_2^k$ is strongly suppressed, indicating extended (ballistic) states in agreement with the collinear model. 
The two descriptions therefore identify the same critical and ballistic regions in twist-angle space, despite their different microscopic implementations.

\begin{figure}[b!]
    \centering
    \includegraphics[width=0.8\columnwidth]{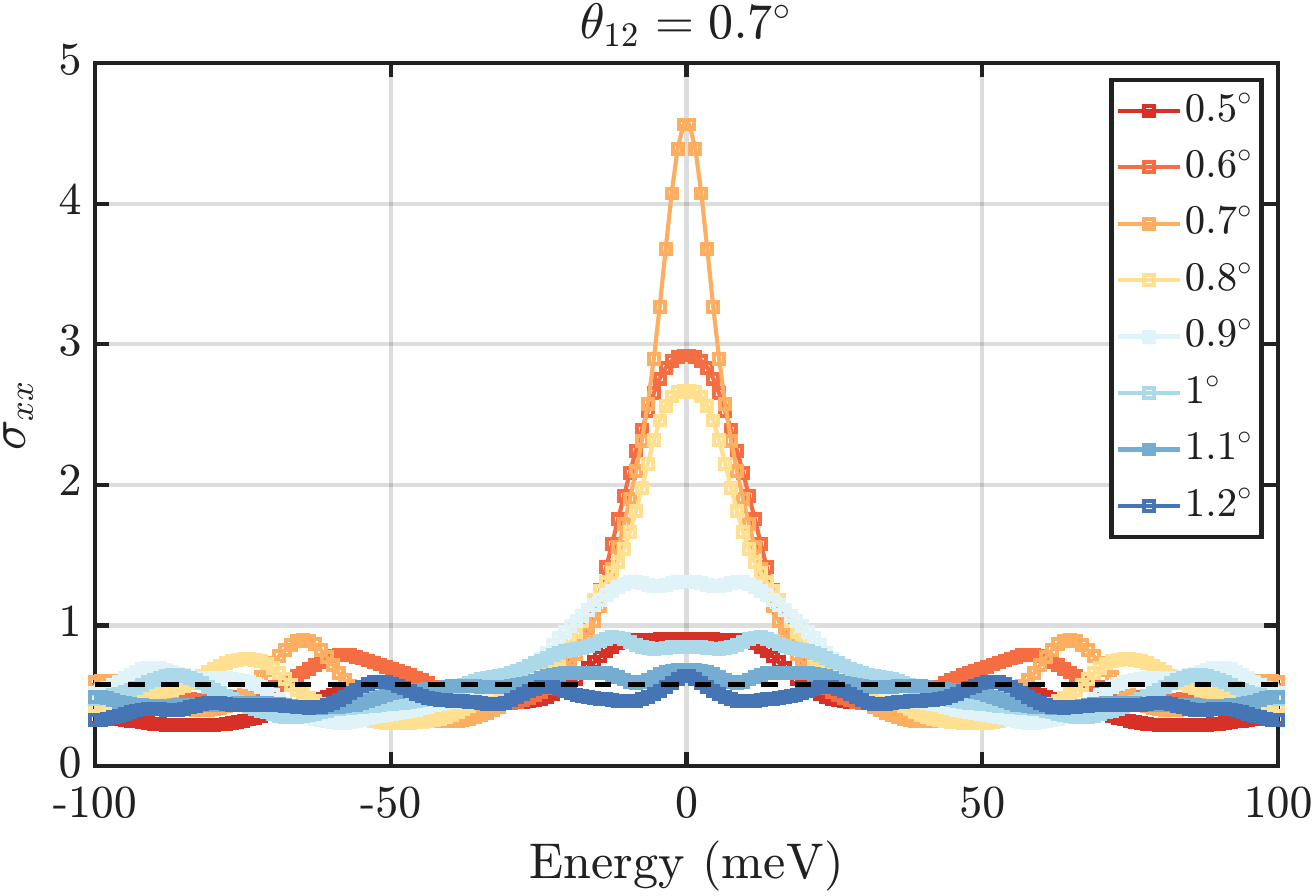}
    \caption{Conductivity in units of $e^2/h$ obtained from the non-collinear model in the chiral limit for $w_\mathrm{AB} = 220$~meV with $\theta_{12}=0.7^\circ$ and varying $\theta_{23}$. The smearing width is 5~meV. The black dashed line is 0.6, which is the IQHT value.}
    \label{fig:sigmaxx_Zoe}
\end{figure}

To further confirm the agreement between the two models, we compare their transport properties. 
\Cref{fig:sigmaxx_Zoe} shows the longitudinal conductivity $\sigma_{xx}(E)$, obtained from the non-collinear model in the chiral limit for $w_\mathrm{AB}=220$~meV at fixed $\theta_{12}=0.7^\circ$ as a function of $\theta_{23}$.
The conductivity is evaluated using the Kubo formula with an energy smearing of $5$~meV, analogous to the KPM calculations in the collinear model. 
Away from commensurate angles, $\sigma_{xx}$ is nearly energy independent and fluctuates around the integer quantum Hall transition value $\sigma_{xx} \approx 0.6\, e^2/h$ (black dashed line), signaling a scale-invariant critical metal. 
In contrast, near commensurate twist-angle combinations, $\sigma_{xx}$ exhibits pronounced peaks and dips, reflecting the crossover to ballistic behavior in periodic or nearly periodic configurations. 
This pattern closely mirrors the conductivity maps and the $N_\mathcal{C}$–scaling diagnostics obtained from the collinear model [\cref{fig:conductivity,fig:sigmaxx-th07-ratio}]. 

These comparisons validate the collinear model for the long-wavelength physics relevant here.
Both models identify the same incommensurate twist angle configurations which realize a critical metal with $\sigma_{xx} \sim e^2/h$, multifractal wave functions, and robust resistance to localization. 
Both also recover ballistic behavior along commensurate lines and in the weakly coupled limits.

\section*{Conclusion}

We have shown that TTG with two generic independent twist angles can realize a quantum-critical metallic phase in its normal state. 
As a consequence, it also hosts robust superconductivity over a broad range of twist-angle combinations and interlayer couplings. 
In 
the normal-state
critical metal, single-particle wave functions are multifractal and the longitudinal conductivity remains of the order of the universal average IQHT value $\sim 0.6e^2/h$ across a wide energy window.

The key insight is the identification of the effective TTG model with that of exotic topological surface states, predicted to demonstrate robust quantum criticality in the \emph{spatially aperiodic structure} of wave functions. 
Our results show that quasiperiodicity, a feature generic in multilayer moir\'e materials, can serve as a source of momentum mixing that reveals the effective topology and enables superconductivity without fine-tuning or degradation of the superfluid stiffness.
Moreover, we show that increasing
the interlayer coupling strength stabilizes quantum criticality in the
normal state and further enhances superconductivity. 
Therefore, we predict that stronger superconductivity occurs in quasiperiodic TTG
if high pressure is applied.

The low energy physics of other moir\'{e} materials, such as transition-metal
dichalcogenide (TMD) multilayers, is not described by Dirac fermions
as in multilayer graphene. Therefore, the specific class-AIII/IQHT-like route to
quantum criticality is not expected to apply to these moir\'{e} materials.
Nonetheless, new kinds of quantum criticality could arise in
quasiperiodic TMD multilayers and it is interesting to clarify whether it favors robust superconductivity.

We have limited ourselves to conventional $s$-wave intervalley pairing
in order to focus on the role of the normal state. 
Strong correlations could
produce exotic nodal pairing, as occurs in the $d$-wave cuprates
and as suggested by recent experiments in moir\'e graphene \cite{oh2021,park2025}.
On one hand, the absence of Anderson's theorem for
non-$s$-wave superconductivity seems to imply strong vulnerability to 
translational symmetry breaking \cite{Ticea2024}. 
At the same time, the strong electronic inhomogeneity
observed in the cuprates \cite{Davis2012}
suggests that the interplay between
nodal pairing, translational symmetry breaking, and strong correlations
involves as-yet undiscovered physics. 
Wedding these elements to the topological mechanisms discussed here is
an important direction for future work.

This work was supported by the Welch Foundation Grant No.~C-1809 
(X.Z.\ and M.S.F.). X.Z.\ also acknowledges support from Paul Scherrer Institute, and 
from the ERC
under the EU’s Horizon 2020 research and innovation program, Grant Agreement 810451.
J.H.W.\ acknowledges support from the National Science Foundation under Grant Number DMR-2238895.
Z.Z.\ acknowledges support from a Stanford Science fellowship and a startup fund at Boston College. 
Z.Z.\ and J.H.W.\ acknowledge Aspen Center for Physics where part of this work was performed, which is supported by a National Science Foundation grant PHY-2210452.

\bibliography{ttg}{}
\bibliographystyle{naturemag}

\end{document}


\title{Supplemental Material: Twisted Trilayer Graphene, Quasiperiodic Superconductor} 
\def\rice{Department of Physics and Astronomy, Rice University, Houston, Texas
77005, USA}
\def\rcqm{Rice Center for Quantum Materials, Rice University, Houston, Texas
77005, USA}
\author{Xinghai Zhang}\affiliation{\rice}
\author{Ziyan Zhu}
\affiliation{Department of Physics, Boston College, 140 Commonwealth Avenue
Chestnut Hill, MA 02467, USA
}
\author{Justin H. Wilson}
\affiliation{Department of Physics and Astronomy, Louisiana State University, Baton Rouge, LA 70803, USA}
\affiliation{Center for Computation and Technology, Louisiana State University, Baton Rouge, LA 70803, USA}
\author{Matthew S. Foster}\affiliation{\rice}\affiliation{\rcqm}
\date{\today}

\maketitle
\tableofcontents


\vspace{0.8cm}

The supplemental material is organized as follows: 
In Sec.~\ref{sec:kpm}, we provide the computational details of the kernel polynomial method in the collinear model. 
In Sec. \ref{sec:conductivity}, we show detailed information on the normal conductivity of helical TTG with varying twist angles, pressure and chiral symmetry breaking term. 
We also compare the conductivity obtained with the collinear model and non-collinear model.
In Sec. \ref{sec:dos}, we show the density of states (DOS) in helical TTG obtained with the collinear model.

\section{Kernel Polynomial Method}~\label{sec:kpm}

In this work, we use the numerically exact kernel polynomial method (KPM) \cite{weibe2006_kpm} to evaluate the superconductivity, superfluid
stiffness, density of states, normal state wave 
functions, and conductivity in the collinear model. In KPM, the physical quantities are expanded in the Chebyshev polynomial basis, and kernels are introduced to suppress the Gibbs oscillation near singularities. 
The key insight is that one can write
\[
\delta(E - E') = \frac1{\pi\sqrt{1-E^2}}\bigg[T_0(E)T_0(E') + 2 \sum_{n=1}^\infty T_n(E)T_n(E')\bigg],
\]
and using this as a base, one can create a whole class of Kernels where the sum is cutoff
\[
K_{N_\mathcal C}(E,E') = \frac1{\pi\sqrt{1-E^2}}\bigg[g_{0,N_\mathcal C} T_0(E)T_0(E') + 2 \sum_{n=1}^{N_\mathcal C} g_{n,N_\mathcal C}T_n(E)T_n(E')\bigg].
\]
Here $T_n(E)=\cos(n\arccos(E))$ is the Chebyshev polynomial, and the $K_{N_\mathcal{C}}$ gives a smooth broadening of the $\delta$-function. In the limit as $N_\mathcal C\rightarrow\infty$ it should converge to $\delta(E-E')$ and one has freedom to choose the $g_{n,N_\mathcal{C}}$ for various properties (like positive definite or to minimize the width); for most purposes the Jackson kernel \cite{weibe2006_kpm} is appropriate and has these properties.
Any function $f(E') = \frac1{\pi\sqrt{1-(E')^2}}[ \mu_0 T_0(E') + 2 \sum_{n=1}^\infty \mu _n T_n(E')]$ can be approximated by a convolution  
\[
f_{N_\mathcal C}(E) = \int_{-1}^1 dE' \, K_{N_\mathcal C}(E, E') f(E') = \frac1{\pi\sqrt{1-E^2}}\bigg[ g_{0,N_\mathcal C} \mu_0 T_0(E) + \sum_{n=1}^{N_\mathcal{C}} g_{n,N_\mathcal C} \mu_n T_n(E)\bigg]
\]
(Note that energies are normalized to be $E \in (-1,1)$ to satisfy the Chebyshev polynomials.)
Since $K_{N_\mathcal C}$ represents a broadened $\delta$-function, we can use it elsewhere, for example in the expression for the conductivity
\begin{align}
    \sigma_{xx}(E) 
    & =
    \frac{2\pi^2}{L^2}
    \mathsf{Tr}
    \left[
        j_x \, \delta(E - H) \, j_x \, \delta(E - H)
    \right] \,. \\
     & \approx 
    \frac{2\pi^2}{L^2}
    \mathsf{Tr}
    \left[
        j_x \, K_{N_\mathcal C}(E,H) \, j_x \, K_{N_\mathcal C}(E,H)
    \right] \,.
\end{align}
The power of this method is that it only involves polynomials of $h_K$ and those polynomials are related to each other by a recurrence relation; thus, the method only involves the action of $H\left|\psi\right>$
and is highly efficient when the Hamiltonian $H$ is sparse, which
enables large-scale calculation compared to diagonalizing the Hamiltonian. In the 
self-consistent BdG calculation for superconductivity,
while the Hamiltonian is no longer sparse, we can still perform the action
of the kinetic term and interaction term on the momentum-space and
real-space wave functions separately. The real-space wave function
can then be obtained from the momentum space one via fast Fourier transform
or vice versa \cite{Zhang2025}. The superconductivity data in Fig.~1 is calculated with KPM on a $77\times77$ lattice.

The normal-state conductivity in Figs.~2 and 3 of the main text are evaluated with KPM implementation of the Kubo formula in Eq.~(6) of the main text. 
We use the standard stochastic evaluation of the trace and perform the calculation on  a $308\times 308$ lattice.
We use the filter and shake method \cite{Zhang2025} to calculate the wave function in Fig.~4{\bf a}, which is performed on 
a
$220\times 220 $ lattice.

\section{Conductivity of Helical Twisted Trilayer Graphene} \label{sec:conductivity}
\subsection{Conductivity in Collinear Model}

\begin{figure}[ht!]
    \centering
    \includegraphics[width=0.9\columnwidth]{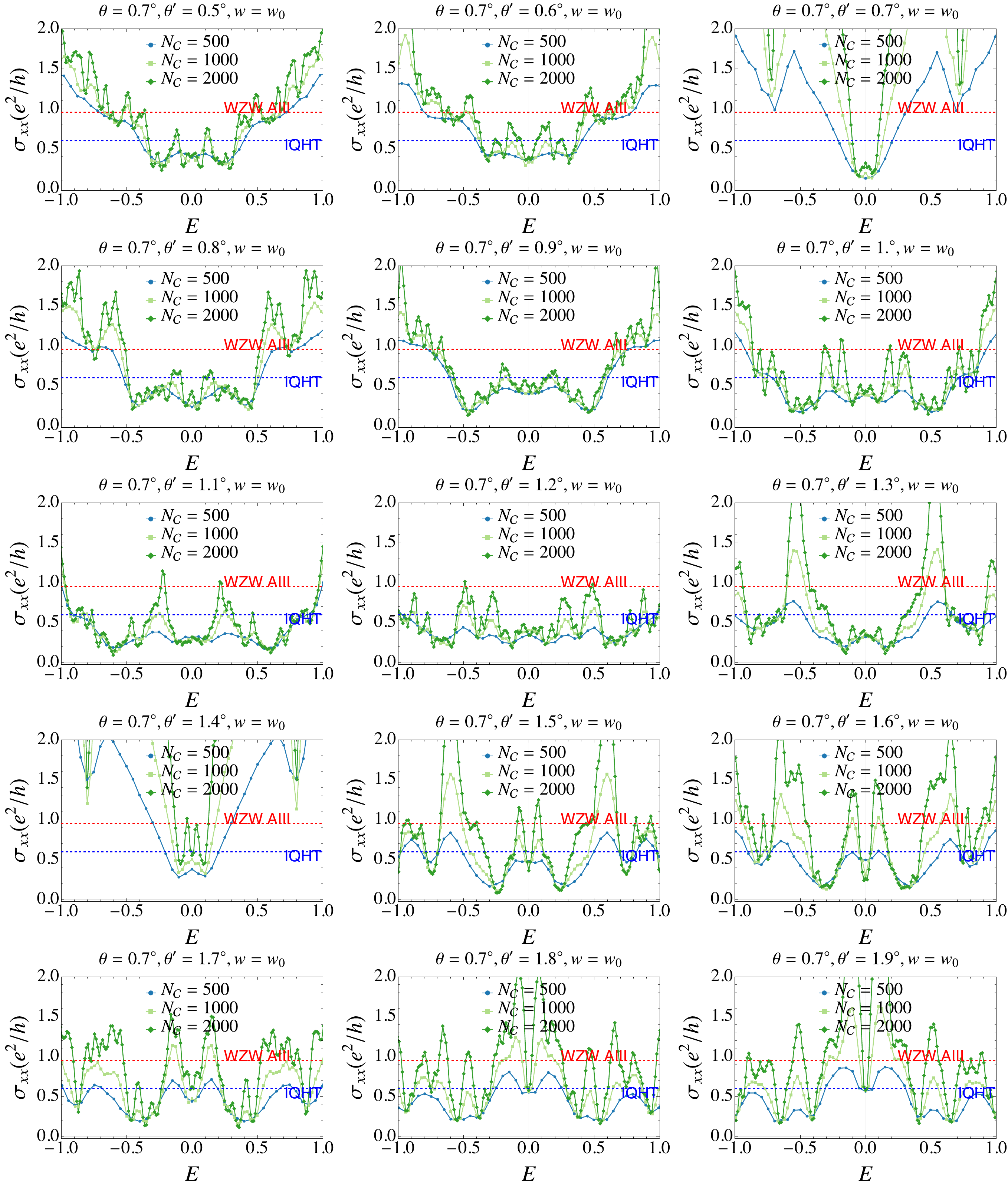}
    \caption{The longitudinal conductivity  in helical twisted trilayer graphene in chiral limit with $\theta=0.7^\circ$, and $w=w_0$.}
    \label{fig:sigmaxx-th07-w1}
\end{figure}

\begin{figure}[ht!]
    \centering
    \includegraphics[width=0.9\columnwidth]{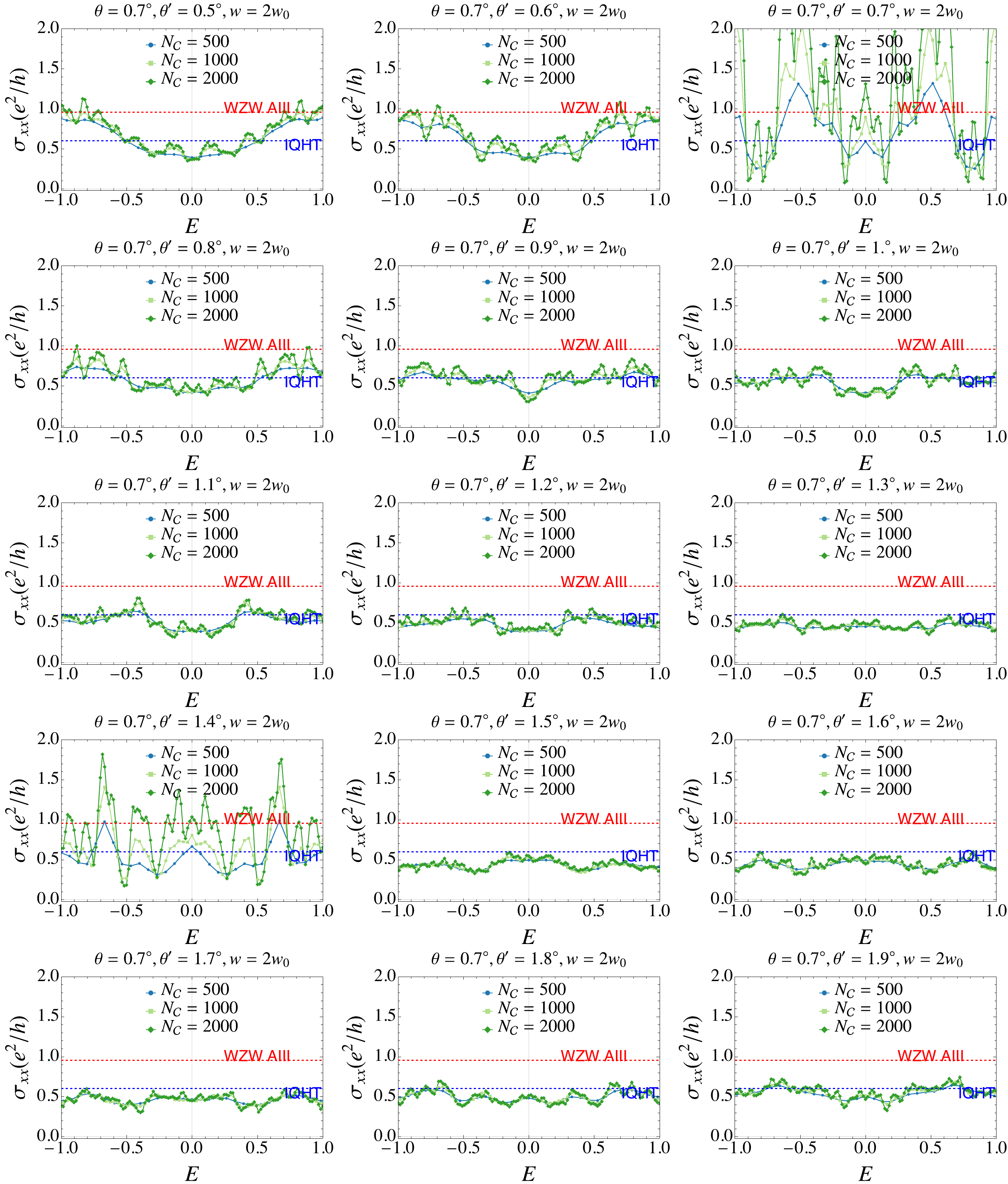}
    \caption{The longitudinal conductivity  in helical twisted trilayer graphene in chiral limit with $\theta=0.7^\circ$, and $w=2w_0$.}
    \label{fig:sigmaxx-th07-w2}
\end{figure}

\begin{figure}[ht!]
    \centering
    \includegraphics[width=0.9\columnwidth]{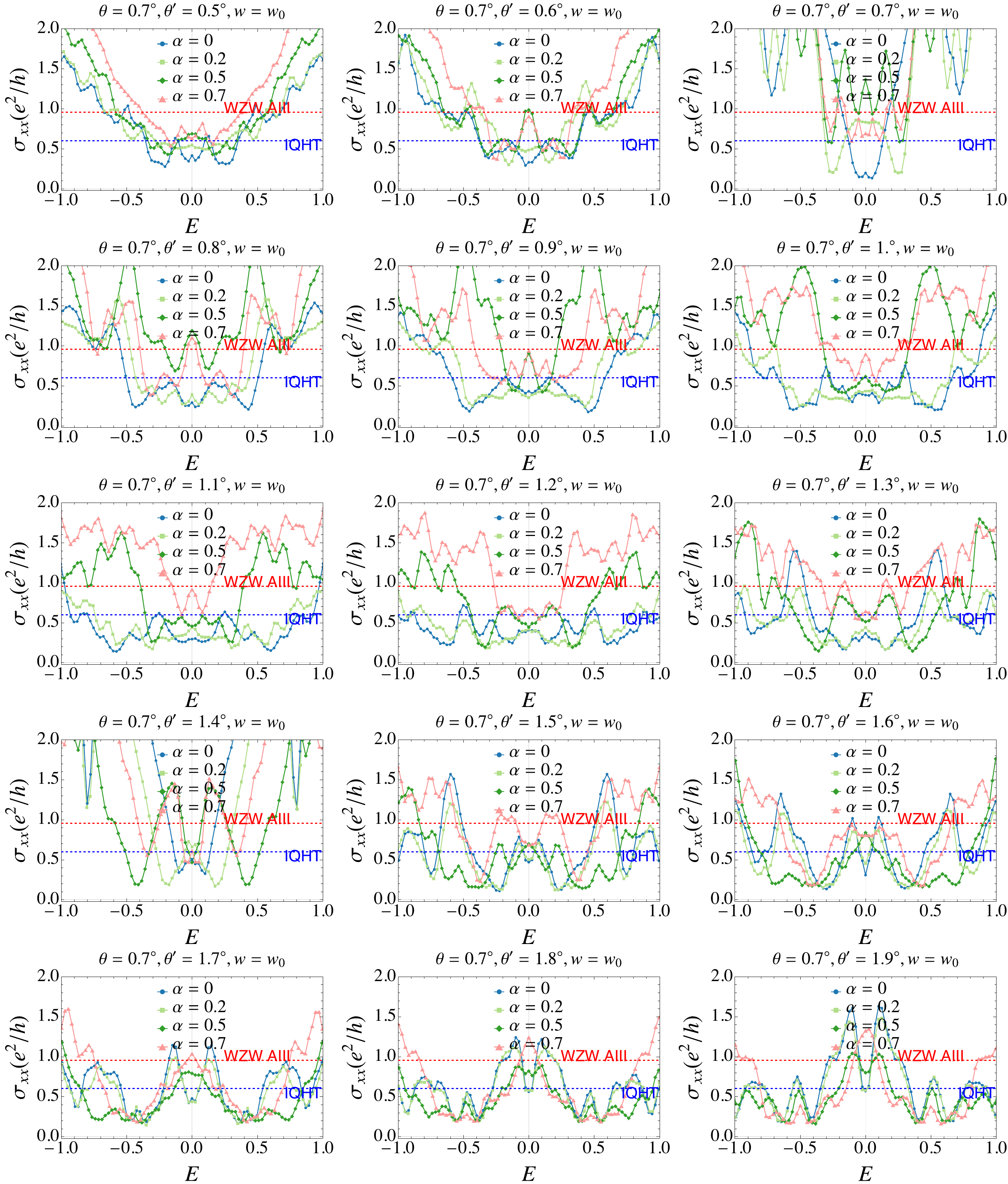}
    \caption{The longitudinal conductivity in helical twisted trilayer graphene with chiral symmetry breaking with $\theta=0.7^\circ$, $\alpha=0.7$ and $w=w_0$.}
    \label{fig:sigmaxx-th07-w1-alpha}
\end{figure}

\begin{figure}[ht!]
    \centering
    \includegraphics[width=0.9\columnwidth]{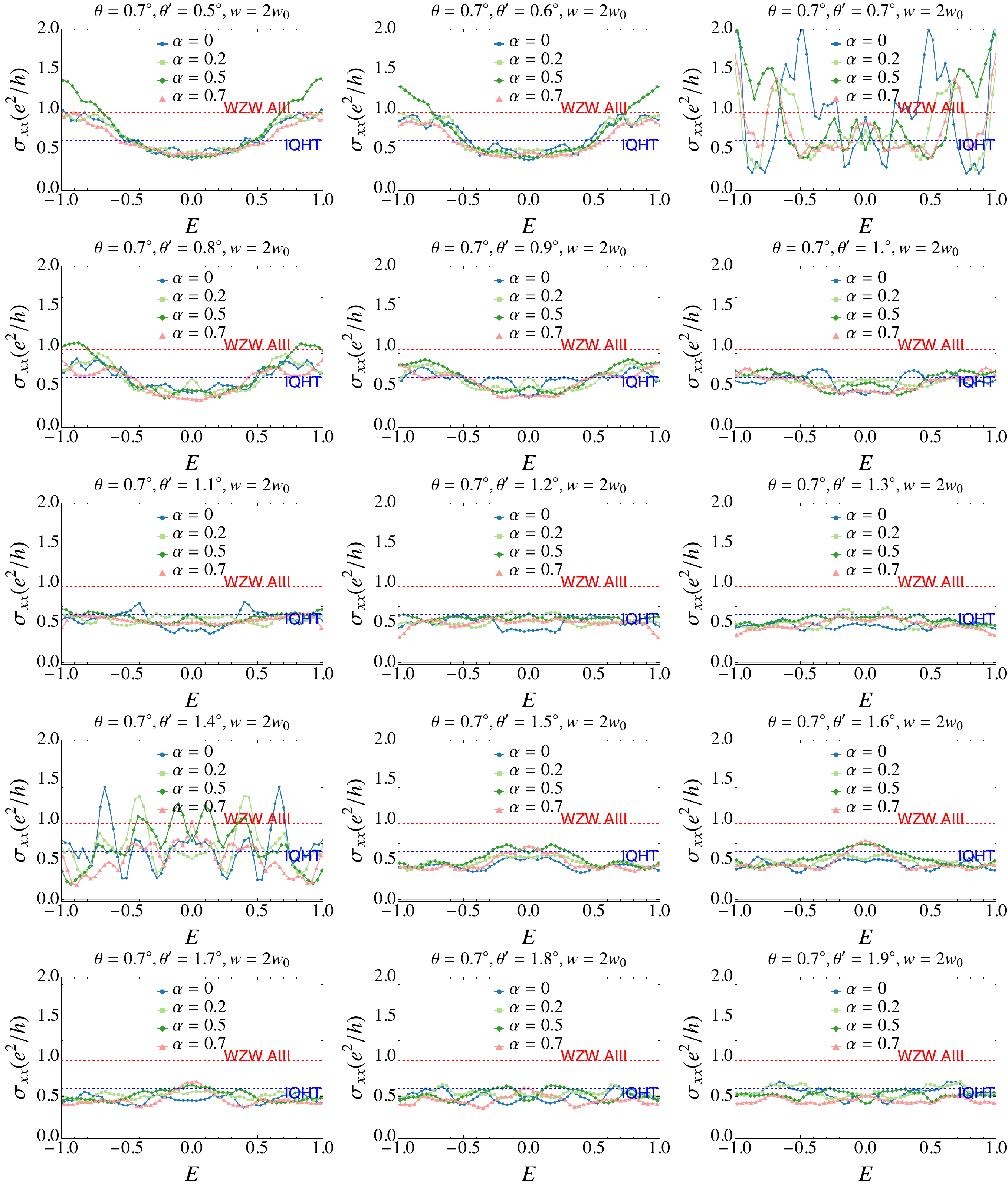}
    \caption{The longitudinal conductivity in helical twisted trilayer graphene with chiral symmetry breaking with $\theta=0.7^\circ$, $\alpha=0.7$ and $w=2w_0$.}
    \label{fig:sigmaxx-th07-w2-alpha}
\end{figure}

Figs.~\ref{fig:sigmaxx-th07-w1} and \ref{fig:sigmaxx-th07-w2} show the normal state conductivity in chiral-limit TTG at ambient and high pressure, respectively. 
The conductivity is calculated with the collinear model of TTG using the KPM implementation of the Kubo formula [Eq. (6) in main text].
At ambient pressure ($w=w_0$), the conductivity shows quantum criticality at incommensurate angles with $\theta'\lesssim 1.2^\circ$, which is reflected by the order of $e^2/h$ value and conductivity convergence with varying order of polynomials $N_\mathcal{C}$.
The non-convergence of $\sigma_{xx}$ at commensurate angles and large $\theta'$ reflects the ballistic transport in the periodic limit.
By increasing the pressure ($w=2w_0$), the quantum criticality is enhanced and $\sigma_{xx}$ shows a constant value in a wide energy window (spectrum-wide quantum criticality, SWQC). 
The value $\sigma_{xx}$ closely follows that of IQHT state ($\sigma_\mathsf{IQHT} \approx 0.6 e^2/h$), which is consistent with the class AIII symmetry of the normal state.

Figs.~\ref{fig:sigmaxx-th07-w1-alpha} and \ref{fig:sigmaxx-th07-w2-alpha} shows the ambient and 
high-pressure conductivity with chiral symmetry breaking ($\alpha \ne 0$). 
At ambient pressure ($w=w_0$), the chiral symmetry breaking term weakens the quantum criticality and the conductivity $\sigma_{xx}$ deviates from that of IQHT state.
However, the quantum criticality is robust against chiral symmetry breaking at high pressure ($w=2w_0$), and the conductivity still follows that of IQHT value in quasiperiodic TTG.

\FloatBarrier

\subsection{Conductivity in Non-Collinear Model}
Here, we test the convergence of the conductivity calculated using the non-collinear model~\cite{Zhu2020} with respect to momentum-space truncation.
In the non-collinear model, interlayer tunneling couples an infinite set of momentum states generated by sums of reciprocal lattice vectors from the three rotated graphene layers.
The basis states are obtained as follows:
\begin{equation}
\bm{k}^{(\ell_3)}
=
\bm{k}^{(\ell_1)}
+
\left(\bm{G}^{(\ell_2)}-\bm{G}^{(\ell_1)}\right)
+
\left(\bm{G}^{(\ell_3)}-\bm{G}'^{(\ell_2)}\right),
\end{equation}
where $\ell_1,\ell_2,\ell_3$ denote cyclic permutations of the layer indices, and
$\bm{G}^{(\ell)}$ ($\bm{G}'^{(\ell)}$) are reciprocal lattice vectors associated with layer $\ell$ in successive interlayer scattering processes.
To truncate this infinite basis, we impose two independent cutoffs:
(i) a radial cutoff $G_{\max}$ restricting the magnitude $|\bm{G}^{(\ell)}|$,
and (ii) a difference cutoff $\Delta G_{\max}$ restricting the momentum transfers
$|\bm{G}^{(\ell)}-\bm{G}^{(\ell')}|$ generated by interlayer tunneling.

\begin{figure}[ht!]
    \centering
    \includegraphics[width=0.8\linewidth]{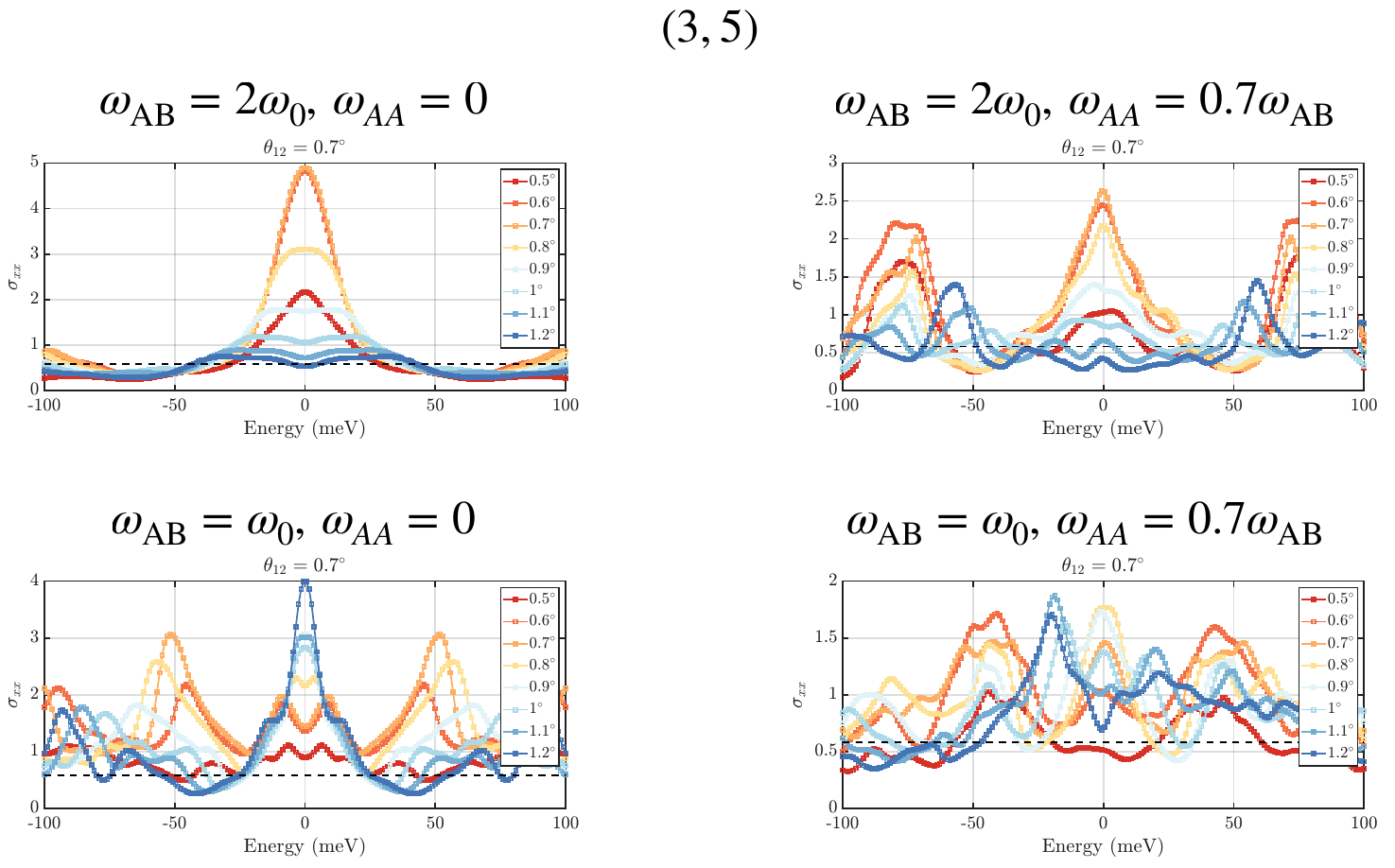}
    \caption{Conductivity in the unit of $e^2/h$ obtained from the non-collinear model in the chiral limit with $\theta_{12}=0.7^\circ$ and varying $\theta_{23}$. The smearing width is 5~meV. The black dashed line is 0.6, which is the IQHT value. Left column: with chiral limit; Right column: no chiral limit, $\omega_{AB} = 0.7 \omega_{AA}$. Top row: with enhanced interlayer coupling $\omega_{AB} = 2\omega_0$; bottom row: without enhanced interlayer coupling $\omega_{AB} = \omega_0$. The cutoff radii are $|\bm{G}^{(\ell)}| < 3 |\bm{G}_1|$ and $|\bm{G}^{(\ell)} - \bm{G}^{(\ell')}| < 5 |\bm{G}_1|$ respectively. }
    \label{fig:sigma_xx_cutoff35}
\end{figure}

\begin{figure}[ht!]
    \centering
    \includegraphics[width=0.8\linewidth]{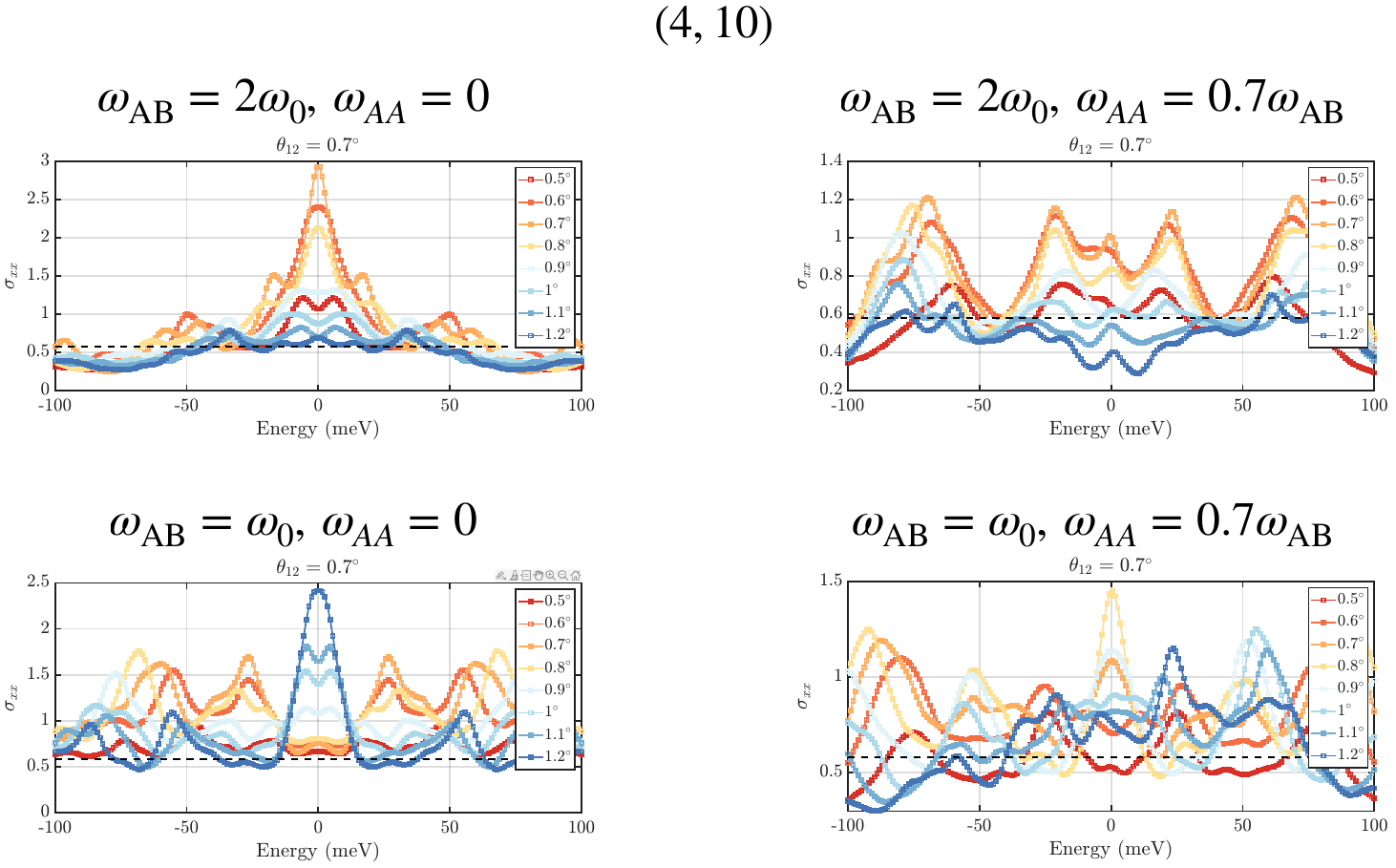}
    \caption{Same as \cref{fig:sigma_xx_cutoff35}, with cutoff radii $|\bm{G}^{(\ell)}| < 4 |\bm{G}_1|$ and $|\bm{G}^{(\ell)} - \bm{G}^{(\ell')}| < 10 |\bm{G}_1|$. }
    \label{fig:sigma_xx_cutoff410}
\end{figure}

\begin{figure}[ht!]
    \centering
    \includegraphics[width=0.8 \linewidth]{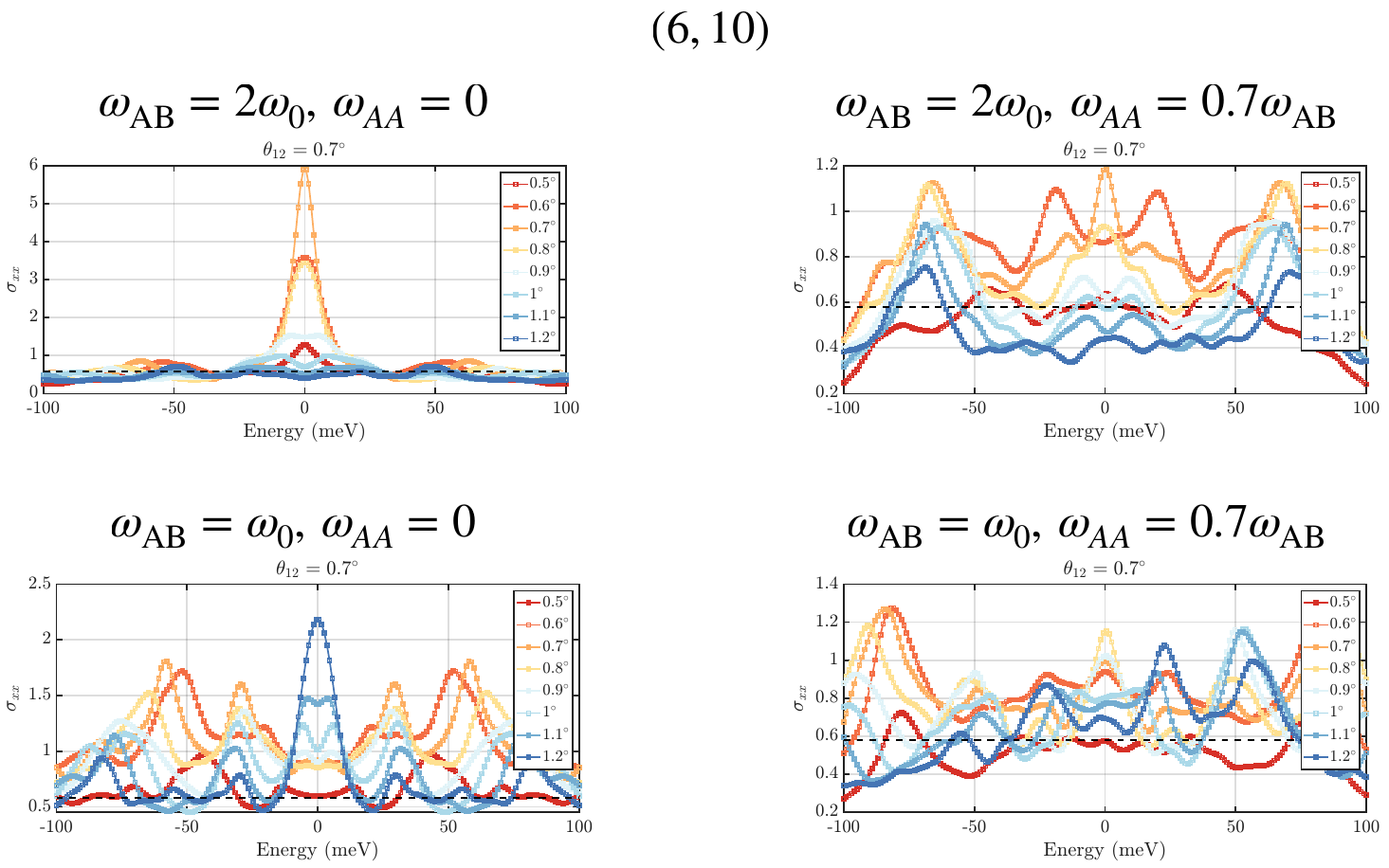}
    \caption{Same as \cref{fig:sigma_xx_cutoff35}, with cutoff radii $|\bm{G}^{(\ell)}| < 6 |\bm{G}_1|$ and $|\bm{G}^{(\ell)} - \bm{G}^{(\ell')}| < 10 |\bm{G}_1|$. }
    \label{fig:sigma_xx_cutoff610}
\end{figure}

In \cref{fig:sigma_xx_cutoff35,fig:sigma_xx_cutoff410,fig:sigma_xx_cutoff610}, we compare the longitudinal conductivity $\sigma_{xx}(E)$ obtained using three representative choices of $(G_{\max},\Delta G_{\max})$,
both in and away from the chiral limit and for both the standard interlayer coupling $w_\mathrm{AB}=w_0$ and the enhanced coupling $w_\mathrm{AB}=2w_0$.
Because the non-collinear model is generically incommensurate, the momentum-space lattice of coupled states is infinite, and strict convergence with a finite cutoff is not guaranteed.
Nevertheless, across all tested cutoffs the qualitative features of the conductivity remain unchanged.
Away from commensurate twist-angle combinations, $\sigma_{xx}(E)$ remains approximately energy independent and of order $e^2/h$,
while near commensurate configurations it exhibits pronounced peaks and dips, reflecting a crossover toward ballistic transport.
Moreover, for the enhanced interlayer coupling $w_\mathrm{AB}=2w_0$, the conductivity is noticeably less sensitive to the choice of cutoff and remains stable across the entire energy window shown,
consistent with stronger momentum mixing and critical behavior discussed in the main text.

\section{Density of States in Helical Twisted Trilayer Graphene}\label{sec:dos}
\begin{figure}[ht!]
    \centering
    \includegraphics[width=0.9\columnwidth]{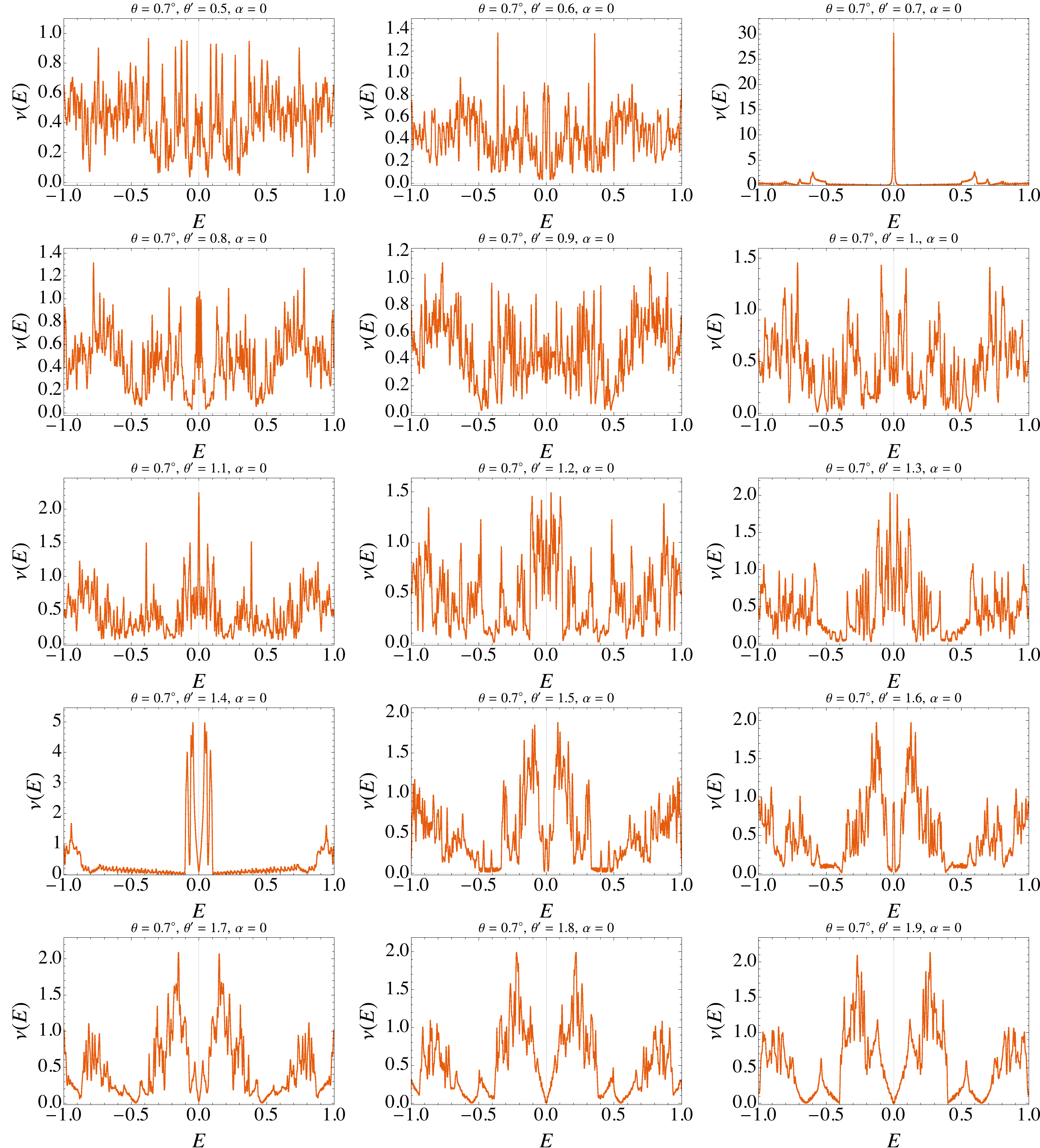}
    \caption{The DOS in helical twisted trilayer graphene at ambient pressure ($w=w_0$) and in the chiral limit ($\alpha=0$).}
    \label{fig:dos-th07-a0}
\end{figure}

\begin{figure}[ht!]
    \centering
    \includegraphics[width=0.9\columnwidth]{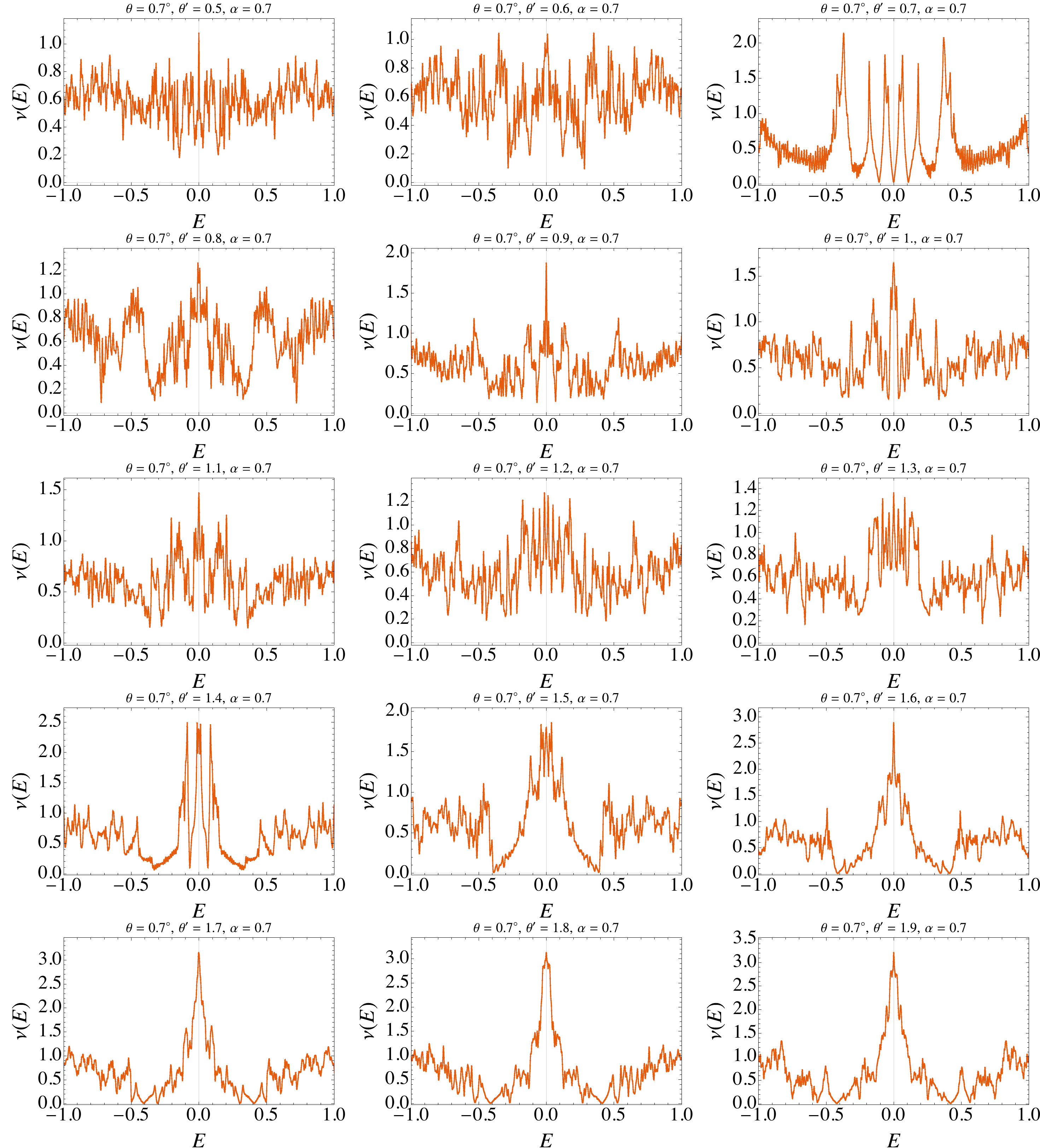}
    \caption{The DOS in helical twisted trilayer graphene  at ambient pressure ($w=w_0$) with chiral symmetry breaking ($\alpha=0.7$).}
    \label{fig:dos-th07-a07}
\end{figure}

Figs.~\ref{fig:dos-th07-a0} and \ref{fig:dos-th07-a07} show the density of states (DOS) of helical twisted trilayer graphene at ambient pressure in the chiral limit and with chiral symmetry breaking, respectively.
At commensurate twist angles, a pronounced zero-energy peak appears in the chiral limit, corresponding to flat bands. When chiral symmetry is broken, these flat bands disappear and the DOS recovers the linear energy dependence characteristic of Dirac fermions.
At incommensurate angles, the DOS exhibits a highly spiky structure across energy, reminiscent of the Aubry–André model, reflecting the quasiperiodic nature of twisted trilayer graphene.

\begin{figure}[ht!]
    \centering
    \includegraphics[width=0.9\columnwidth]{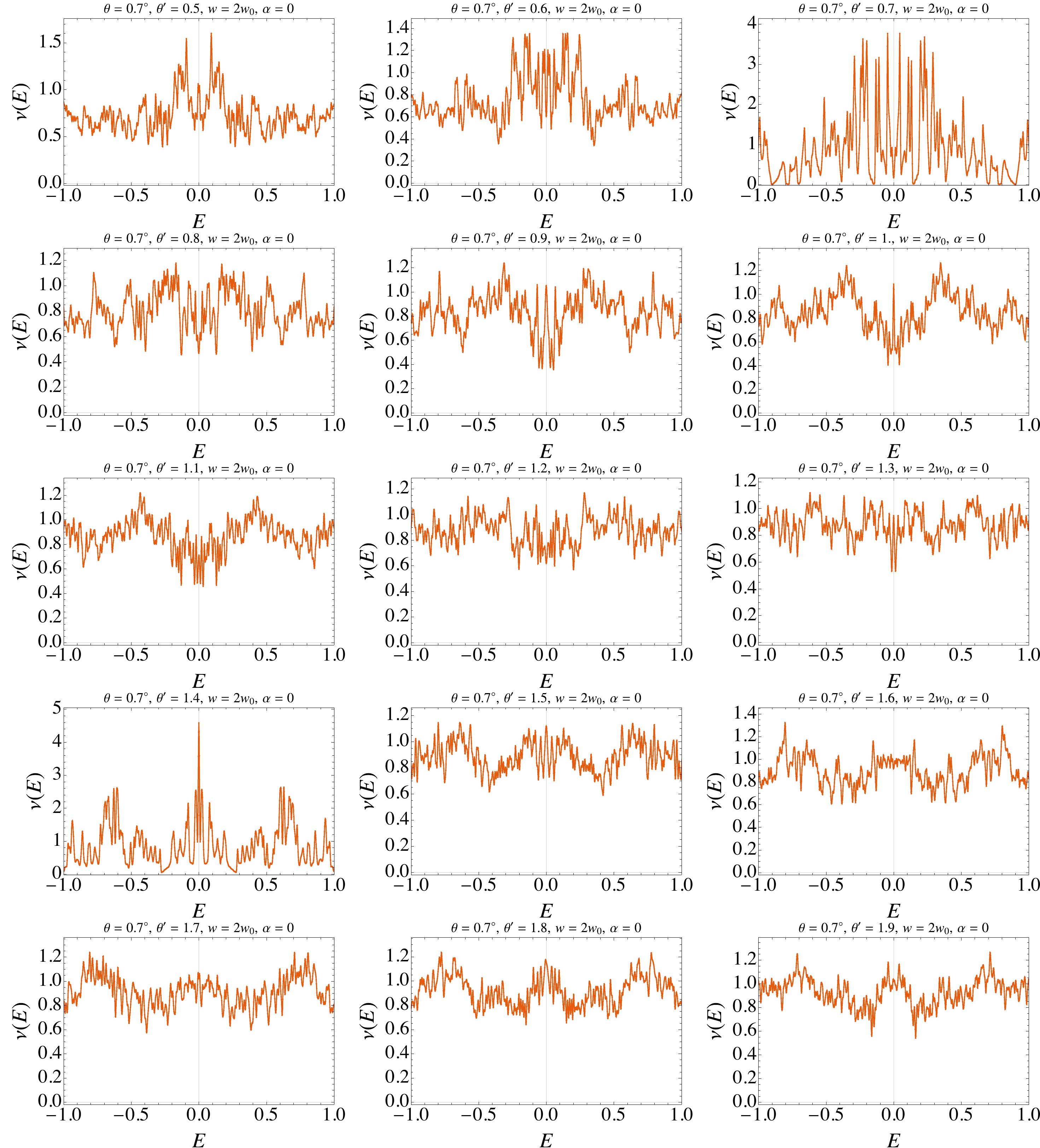}
    \caption{The DOS in helical twisted trilayer graphene in the chiral limit ($\alpha=0$) and at high pressure ($w=2w_0$).}
    \label{fig:dos-th07-a0-w2}
\end{figure}

\begin{figure}[ht!]
    \centering
    \includegraphics[width=0.9\columnwidth]{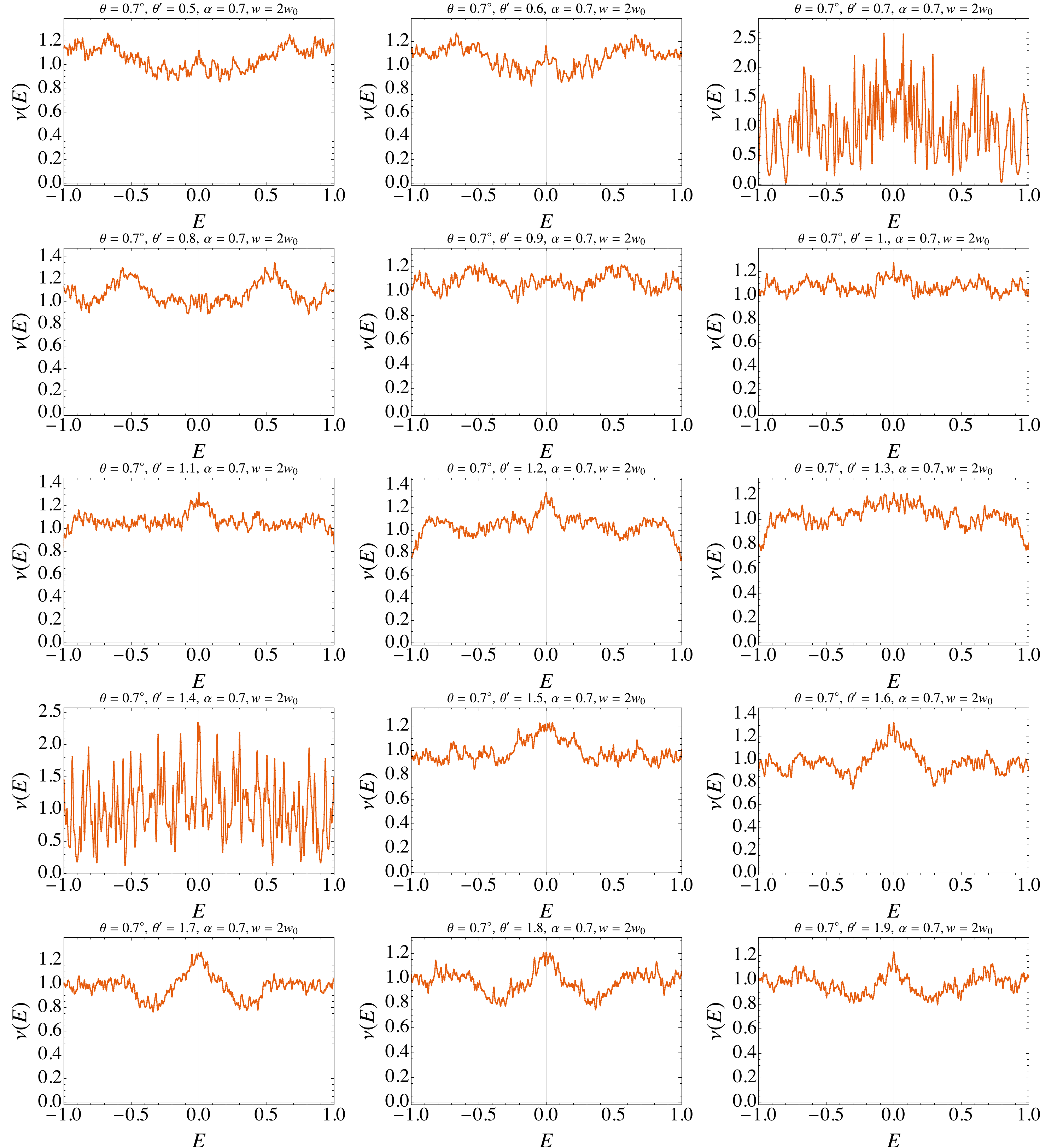}
    \caption{The DOS in helical twisted trilayer graphene with chiral symmetry breaking term ($\alpha=0.7$) and at high pressure ( $w=2w_0$).}
    \label{fig:dos-th07-a07-w2}
\end{figure}

Figs.~\ref{fig:dos-th07-a0-w2} and \ref{fig:dos-th07-a07-w2} show the DOS at higher pressure ($w = 2w_0$).
Compared with ambient pressure, the DOS is significantly smoother, consistent with the enhancement of quantum criticality at enhanced pressure. 


\FloatBarrier

\bibliography{ttgSM}{}
\bibliographystyle{naturemag}